\documentclass[twocolumn, tighten, twocolappendix]{aastex7}
\usepackage{color} 
\usepackage{multirow}

\def\ltsim{\mathrel{\hbox{\rlap{\hbox{\lower4pt\hbox{$\sim$}}}\hbox{$<$}}}}
\def\gtsim{\mathrel{\hbox{\rlap{\hbox{\lower4pt\hbox{$\sim$}}}\hbox{$>$}}}}

\newcommand\degrees{\mbox{$^{\circ}$}}

\hypersetup{pdfauthor={The UNIONS Collaboration},
            pdftitle={},
            pdfkeywords={},
            bookmarksnumbered=true}

\begin{document}
\shorttitle{UNIONS}
\shortauthors{The UNIONS Collaboration}
\title{UNIONS: The Ultraviolet Near-Infrared Optical Northern Survey}
\author[0000-0001-8221-8406]{Stephen Gwyn}
\affiliation{National Research Council Herzberg Astronomy and Astrophysics, 5071 West Saanich Road, Victoria, B.C., V8Z6M7, Canada}
\email{stephen.gwyn@nrc-cnrc.gc.ca}

\author[0000-0003-4666-6564]{Alan W. McConnachie}
\affiliation{National Research Council Herzberg Astronomy and Astrophysics, 5071 West Saanich Road, Victoria, B.C., V8Z6M7, Canada}
\email{alan.mcConnachie@nrc-cnrc.gc.ca}

\author[0000-0002-3263-8645 ]{Jean-Charles Cuillandre}
\affiliation{Universit\'e Paris-Saclay, Universit\'e Paris Cit\'e, CEA, CNRS, AIM, 91191, Gif-sur-Yvette, France }
\email{jcc@cfht.hawaii.edu}

\author[0000-0001-6965-7789 ]{Ken C. Chambers}
\affiliation{Institute for Astronomy, University of Hawai'i, 2680 Woodlawn Drive, Honolulu, HI 96822, USA}
\email{chambers@ifa.hawaii.edu}

\author[0000-0002-7965-2815 ]{Eugene A. Magnier}
\affiliation{Institute for Astronomy, University of Hawai'i, 2680 Woodlawn Drive, Honolulu, HI 96822, USA}
\email{eugene@ifa.hawaii.edu}

\author[0000-0002-1437-3786 ]{Michael J. Hudson}
\affiliation{Department of Physics and Astronomy, University of Waterloo, 200 University Avenue West, Waterloo, Ontario N2L 3G1, Canada}\affiliation{Waterloo Centre for Astrophysics, University of Waterloo, Waterloo, Ontario N2L 3G1, Canada}\affiliation{Perimeter Institute for Theoretical Physics, 31 Caroline St. North, Waterloo, ON N2L 2Y5, Canada}
\email{mike.hudson@uwaterloo.ca}

\author[0000-0003-3484-399X]{Masamune Oguri}
\affiliation{Center for Frontier Science, Chiba University, 1-33 Yayoicho, Inage, Chiba 263-8522, Japan}\affiliation{Department of Physics, Graduate School of Science, Chiba University, 1-33 Yayoicho, Inage, Chiba 263-8522, Japan}
\email{masamune.oguri@chiba-u.jp}

\author[0000-0002-6174-8165]{Hisanori Furusawa}
\affiliation{Astronomy Data Center, National Astronomical Observatory of Japan, 2-21-1 Osawa, Mitaka, Tokyo 181-8588, Japan}\affiliation{The Graduate University for Advanced Studies, SOKENDAI, 2-21-1 Osawa, Mitaka, Tokyo 181-8588, Japan}
\email{furusawa.hisanori@nao.ac.jp}

\author[0000-0002-9814-3338]{Hendrik Hildebrandt}
\affiliation{Ruhr University Bochum, Faculty of Physics and Astronomy, Astronomical Institute (AIRUB), German Centre for Cosmological Lensing, 44780 Bochum, Germany}
\email{hendrik@astro.ruhr-uni-bochum.de}

\author[0000-0002-7667-0081]{Raymond Carlberg}
\affiliation{Department of Astronomy and Astrophysics, University of Toronto, 50 St George St, Toronto, ON M5S 3H4, Canada }
\email{raymond.carlberg@utoronto.ca}

\author[0000-0002-1768-1899]{Sara L. Ellison}
\affiliation{Department of Physics \& Astronomy, University of Victoria, Finnerty Road, Victoria, BC V8P 1A1, Canada}
\email{sarae@uvic.ca}

\author[0000-0002-1968-5762]{Junko Furusawa}
\affiliation{National Astronomical Observatory of Japan, 2-21-1 Osawa, Mitaka, Tokyo 181-8588, Japan}
\email{furusawa.junko@nao.ac.jp}

\author[0000-0002-5540-6935]{Rapha\"el  Gavazzi}
\affiliation{Laboratoire d'Astrophysique de Marseille, CNRS, Aix-Marseille Universit\'e, CNES, Marseille, France} \affiliation{Institut d'Astrophysique de Paris, UMR 7095, CNRS, and Sorbonne Universit\'e, 98 bis boulevard Arago, 75014 Paris} 
\email{raphael.gavazzi@lam.fr}

\author[0000-0002-3292-9709]{Rodrigo Ibata}
\affiliation{Universit\'e de Strasbourg, CNRS, Observatoire astronomique de Strasbourg, UMR 7550, F-67000 Strasbourg, France}
\email{rodrigo.ibata@astro.unistra.fr}

\author{Yannick Mellier}
\affiliation{Institut d'Astrophysique de Paris, 98bis Boulevard Arago, 75014, Paris, France}
\email{yannick.paul.mellier@gmail.com}

\author[0000-0002-7934-2569]{Ken Osato}
\affiliation{Center for Frontier Science, Chiba University, Chiba 263-8522, Japan}
\email{ken.osato@chiba-u.jp}


\author[0000-0002-1371-5705]{H. Aussel}
\affiliation{Universit\'e Paris-Saclay, Universit\'e Paris Cit\'e, CEA, CNRS, AIM, 91191, Gif-sur-Yvette, France}
\email{herve.aussel@cea.fr}

\author[0000-0002-1518-0150]{Lucie Baumont}
\affiliation{Dipartimento di Fisica - Sezione di Astronomia, Università di Trieste, Via Tiepolo 11, 34131 Trieste, Italy}\affiliation{INAF-Osservatorio Astronomico di Trieste, Via G.~B.~Tiepolo 11, 34143 Trieste, Italy}
\email{lucie.baumont@gmail.com}

\author[0000-0002-8068-0645]{Manuel Bayer}
\affiliation{Kapteyn Astronomical Institute, University of Groningen, Landleven 12, NL-9747AD Groningen, the Netherlands}
\email{mbayer@astro.rug.nl}

\author{Olivier Boulade}
\affiliation{Astrophysics Department, Centre d'Etudes de Saclay, 91191 Gif sur Yvette FRANCE}
\email{ olivier.boulade@cea.fr}

\author[0000-0003-1184-8114]{Patrick C\^ot\'e}
\affiliation{National Research Council Herzberg Astronomy and Astrophysics, 5071 West Saanich Road, Victoria, B.C., V8Z6M7, Canada}
\email{Patrick.Cote@nrc-cnrc.gc.ca}

\author[0009-0001-4503-3071]{David Chemaly}
\affiliation{Institute of Astronomy, University of Cambridge, Madingley Road, Cambridge CB3 0HA, UK}
\email{dc824@cam.ac.uk}

\author[0000-0002-3760-2086]{Cail Daley}
\affiliation{Universit\'e Paris-Saclay, Universit\'e Paris Cit\'e, CEA, CNRS, AIM, 91191, Gif-sur-Yvette, France}
\email{cail.daley@cea.fr}

\author[0000-0003-3343-6284]{Pierre-Alain Duc}
\affiliation{Universit\'e de Strasbourg, CNRS, Observatoire astronomique de Strasbourg, UMR 7550, F-67000 Strasbourg, France}
\email{pierre-alain.duc@astro.unistra.fr}

\author[0000-0002-1038-3370]{A. Ellien}
\affiliation{OCA, P.H.C Boulevard de l'Observatoire CS 34229, 06304 Nice Cedex 4, France}
\email{amael.ellien@oca.eu}

\author[0000-0003-2239-7988]{S\'ebastien Fabbro}
\affiliation{National Research Council Herzberg Astronomy and Astrophysics, 5071 West Saanich Road, Victoria, B.C., V8Z6M7, Canada}
\email{sebastien.fabbro@nrc-cnrc.gc.ca}

\author[0000-0002-8919-079X]{Leonardo Ferreira}
\affiliation{Department of Physics \& Astronomy, University of Victoria, Finnerty Road, Victoria, BC V8P 1A1, Canada} \affiliation{Instituto de Matemática Estatística e Física, Universidade Federal do Rio Grande, Rio Grande, RS, Brazil}
\email{leonardo.ferreira.furg@gmail.com}

\author[0000-0002-5956-8018]{Itsna K. Fitriana}
\affiliation{National Astronomical Observatory of Japan, 2-21-1 Osawa, Mitaka, Tokyo 181-8588, Japan} \affiliation{Department of Astronomy, Institut Teknologi Bandung, Jl. Ganesha 10, Bandung 40132, Indonesia} 
\email{itsna.fitriana@nao.ac.jp}

\author{Emeric Le Floc'h}
\affiliation{Universit\'e Paris-Saclay, Universit\'e Paris Cit\'e, CEA, CNRS, AIM, 91191, Gif-sur-Yvette, France}
\email{emeric.lefloch@cea.fr}

\author[0000-0002-2165-5044]{Hammer, Francois}
\affiliation{LIRA, Paris Observatory - PSL, CNRS, 77 Av. Denfert Rochereau, 75014 Paris, France}
\email{francois.hammer@obspm.fr}

\author[0000-0001-7440-8832]{Yoshinobu Fudamoto}
\affiliation{Center for Frontier Science, Chiba University, 1-33 Yayoi-cho, Inage-ku, Chiba 263-8522, Japan} \affiliation{Steward Observatory, University of Arizona, 933 N Cherry Avenue, Tucson, AZ 85721, USA}
\email{yoshinobu.fudamoto@gmail.com}

\author[0000-0003-1015-5367]{Hua Gao}
\affiliation{Institute for Astronomy, University of Hawaii, 2680 Woodlawn Drive, Honolulu, HI 96822, USA}
\email{hgao@hawaii.edu}

\author[0000-0002-0104-8132]{L. W. K. Goh}
\affiliation{Universit\'{e} Paris-Saclay, Universit\'{e} Paris Cit\'{e}, CEA, CNRS, AIM}\affiliation{Institute for Astronomy, University of Edinburgh, Royal Observatory, Blackford Hill, Edinburgh EH9 3HJ, U.K.}
\email{lgoh@roe.ac.uk}

\author[0000-0002-6821-8669]{Tomotsugu Goto}
\affiliation{Institute of Astronomy, National Tsing Hua University, 101, Section 2, Kuang-Fu Road, Hsinchu 30013, Taiwan (R.O.C.)}
\email{tomo@gapp.nthu.edu.tw}

\author[0009-0004-3655-4870]{Sacha Guerrini}
\affiliation{Universit\'e Paris Cit\'e, Universit\'e Paris-Saclay, CEA, CNRS, AIM, F-91191, Gif-sur-Yvette, France}
\email{sacha.guerrini@cea.fr}

\author[0000-0002-5068-7918]{Axel Guinot}
\affiliation{Department of Physics, McWilliams Center for Cosmology, Carnegie Mellon University, Pittsburgh, PA 15213, USA}
\email{axel.guinot.astro@gmail.com}

\author[0000-0003-2927-5465]{Vincent H\'{e}nault-Brunet}
\affiliation{Department of Astronomy and Physics, Saint Mary's University, 923 Robie Street, Halifax, B3H 3C3, Canada}
\email{Vincent.Henault@smu.ca}

\author[0000-0002-6047-430X]{Yuichi Harikane}
\affiliation{Institute for Cosmic Ray Research, The University of Tokyo, 5-1-5 Kashiwanoha, Kashiwa, Chiba 277-8582, Japan}
\email{hari@icrr.u-tokyo.ac.jp}

\author[0000-0002-8758-8139]{Kohei Hayashi}
\affiliation{National Institute of Technology, Sendai College, Natori, Miyagi 981-1239, Japan} \affiliation{Astronomical Institute, Tohoku University, Aoba-ku, Sendai 980-8578, Japan} \affiliation{ICRR, The University of Tokyo, Kashiwa, Chiba 277-8582, Japan}
\email{khayashi@sendai-nct.ac.jp}

\author[0000-0003-1196-4940]{Nick Heesters}
\affiliation{Institute of Physics, Laboratory of Astrophysics, Ecole Polytechnique F\'ed\'erale de Lausanne (EPFL), 1290 Sauverny, Switzerland}
\email{nick.heesters@epfl.ch}

\author[0000-0002-4377-903X]{Kohei Ichikawa}
\affiliation{Waseda University, 1 Chome-104 Totsukamachi, Shinjuku City, Tokyo 169-8050, Japan}
\email{ichikawa.waseda@gmail.com}

\author[0000-0001-9513-7138]{Martin Kilbinger}
\affiliation{Universit\'e Paris-Saclay, Universit\'e Paris Cit\'e, CEA, CNRS, AIM, 91191, Gif-sur-Yvette, France}
\email{martin.kilbinger@cea.fr}

\author[0000-0003-1980-8838]{P. B. Kuzma}
\affiliation{National Astronomical Observatory of Japan, 2-21-1 Osawa, Mitaka, Tokyo 181-8588, Japan}
\email{pete.kuzma@nao.ac.jp}

\author[0000-0003-3616-6486]{Qinxun Li}
\affiliation{Department of Physics and Astronomy, University of Utah, Salt Lake City, UT 84102, USA}
\email{qinxun.li@utah.edu}

\author[0000-0002-9104-314X]{Tob{\'i}as I. Liaudat}
\affiliation{IRFU, CEA, Universit{\'e} Paris-Saclay F-91191 Gif-sur-Yvette Cedex, France}
\email{tobias.liaudat@cea.fr}

\author[0000-0002-7272-5129]{Chien-Cheng Lin}
\affiliation{Institute for Astronomy, University of Hawai'i, 2680 Woodlawn Drive, Honolulu, HI 96822, USA}
\email{cclin33@hawaii.edu}

\author[0000-0003-4552-9808]{Oliver M\"uller}
\affiliation{Institute of Physics, Laboratory of Astrophysics, Ecole Polytechnique F\'ed\'erale de Lausanne (EPFL), 1290 Sauverny, Switzerland}\affiliation{Institute of Astronomy, University of Cambridge, Madingley Road, Cambridge CB3 0HA, UK}
\email{oliver.muller@epfl.ch}

\author[0000-0002-1349-202X]{Nicolas F. Martin}
\affiliation{Universit\'e de Strasbourg, CNRS, Observatoire astronomique de Strasbourg, UMR 7550, F-67000 Strasbourg, France} \affiliation{Max-Planck-Institut f\"{u}r Astronomie, K\"{o}nigstuhl 17, D-69117 Heidelberg, Germany}
\email{nicolas.martin@astro.unistra.fr}

\author[0000-0001-5063-0340]{Yoshiki Matsuoka}
\affiliation{Research Center for Space and Cosmic Evolution, Ehime University, 3 Bunkyo-cho, Matsuyama-shi, Ehime, Japan}
\email{ yk.matsuoka@cosmos.ehime-u.ac.jp}

\author[0000-0003-0105-9576]{Gustavo E. Medina}
\affiliation{David A. Dunlap Department of Astronomy \& Astrophysics, University of Toronto, 50 St George Street, Toronto ON M5S 3H4, Canada}\affiliation{Dunlap Institute for Astronomy \& Astrophysics, University of Toronto, 50 St George Street, Toronto, ON M5S 3H4, Canada}
\email{gustavo.medina@utoronto.ca}

\author[0000-0001-7964-9766]{Hironao Miyatake}
\affiliation{Kobayashi-Maskawa Institute for the Origin of Particles and the Universe (KMI), Nagoya University, Nagoya, 464-8602, Japan Institute for Advanced Research, Nagoya University, Nagoya, 464-8601, Japan Kavli Institute for the Physics and Mathematics of the Universe (WPI), The University of Tokyo Institutes for Advanced Study (UTIAS), The University of Tokyo, Chiba 277-8583, Japan}
\email{hironao.miyatake@nagoya-u.jp}

\author[0000-0002-1962-904X]{Satoshi Miyazaki}
\affiliation{Subaru Telescope, National Astronomical Observatory of Japan, 2-21-1 Osawa, Mitaka, Tokyo 181-8588, Japan}
\email{satoshi@naoj.org}

\author[0000-0002-7805-2500]{Charlie T. Mpetha}
\affiliation{Institute for Astronomy, School of Physics and Astronomy, University of Edinburgh, Royal Observatory, Blackford Hill, Edinburgh, EH9 3HJ, United Kingdom}
\email{c.mpetha@ed.ac.uk}

\author[0000-0002-7402-5441]{Tohru Nagao}
\affiliation{Research Center for Space and Cosmic Evolution, Ehime University, , 3 Bunkyo-cho, Matsuyama-shi, Ehime, Japan}
\email{tohru@cosmos.ehime-u.ac.jp}

\author{Julio F. Navarro}
\affiliation{Department of Physics \& Astronomy, University of Victoria, Finnerty Road, Victoria, BC V8P 1A1, Canada}
\email{jfn@uvic.ca}

\author[0000-0003-3102-7452]{Masafumi Niwano}
\affiliation{National Astronomical Observatory of Japan, 2-21-1 Osawa, Mitaka, Tokyo 181-8588, Japan}
\email{masafumi.niwano@nao.ac.jp}

\author[0000-0001-8239-4549]{Itsuki Ogami}
\affiliation{Department of Astronomy, School of Science, The Graduate University for Advanced Studies (SOKENDAI), 2-21-1 Osawa, Mitaka, Tokyo 181-8588, Japan} \affiliation{National Astronomical Observatory of Japan, 2-21-1 Osawa, Mitaka, Tokyo 181-8588, Japan}
\email{itsuki.ogami@grad.nao.ac.jp}

\author[0000-0003-2898-0728]{Nobuhiro Okabe}
\affiliation{Physics Program, Graduate School of Advanced Science and Engineering, Hiroshima University, 1-3-1 Kagamiyama, Higashi-Hiroshima, Hiroshima 739-8526, Japan}
\email{okabe@hiroshima-u.ac.jp}

\author[0000-0003-2984-6803]{Masafusa Onoue}
\affiliation{Kavli Institute for the Physics and Mathematics of the Universe (Kavli IPMU, WPI), The University of Tokyo Institutes for Advanced Study, The University of Tokyo, Kashiwa, Chiba 277-8583, Japan}
\email{monoue@icloud.com}

\author[0000-0002-6639-6533]{Gregory S.H. Paek}
\affiliation{Institute for Astronomy, University of Hawaii, 2680 Woodlawn Drive, Honolulu, HI 96822, USA}
\email{gregorypaek94@gmail.com}

\author[0000-0003-4722-5744]{Laura C. Parker}
\affiliation{Department of Physics and Astronomy, McMaster University, Hamilton ON L8S 4M1, Canada}
\email{lparker@mcmaster.ca}

\author[0000-0002-1871-4154]{David R. Patton}
\affiliation{Department of Physics and Astronomy, Trent University, 1600 West Bank Drive, Peterborough, ON, K9L 0G2, Canada}
\email{dpatton@trentu.ca}

\author{Fabian Hervas Peters}
\affiliation{Universit\'{e} Paris-Saclay, Universit\'{e} Paris Cit\'{e}, CEA, CNRS, AIM}
\email{fabi.hetvas@gmail.com}

\author[0000-0002-1755-4582]{Simon Prunet}
\affiliation{Universit\'e C\^ote d’Azur, Observatoire de la C\^ote d’Azur, CNRS, Laboratoire Lagrange, Bd de l’Observatoire, CS 34229, 06304 Nice cedex 4, France}
\email{simon.prunet@oca.eu}

\author[0000-0003-4945-0056]{Rub\'en S\'anchez-Janssen}
\affiliation{UK Astronomy Technology Centre, Royal Observatory, Blackford Hill, Edinburgh EH9 3HJ, UK}
\email{ruben.sanchez-janssen@stfc.ac.uk}

\author{M. Schultheis}
\affiliation{Universit\'e C\^ote d’Azur, Observatoire de la C\^ote d’Azur, Laboratoire Lagrange, CNRS, Blvd de l’Observatoire, 06304 Nice, France}
\email{mathias.schultheis@oca.eu}

\author[0000-0002-3182-3574]{Federico Sestito}
\affiliation{University of Hertfordshire, Hatfield, Hertfordshire, AL10 9AB, UK}
\email{f.sestito@herts.ac.uk}

\author[0000-0002-6946-8280]{Simon E.\,T. Smith}
\affiliation{Department of Physics \& Astronomy, University of Victoria, Finnerty Road, Victoria, BC V8P 1A1, Canada}
\email{simonsmith@uvic.ca}

\author[0000-0003-2177-7794]{J.-L. Starck}
\affiliation{Universit\'e Paris-Saclay, Universit\'e Paris Cit\'e, CEA, CNRS, AIM, 91191, Gif-sur-Yvette, France}\affiliation{Institute of Computer Science, Foundation for Research and Technology-Hellas (FORTH), Heraklion, 70013, Greece}
\email{jstarck@cea.fr}

\author[0000-0003-4501-103X]{Else Starkenburg}
\affiliation{Kapteyn Astronomical Institute, University of Groningen, Landleven 12, NL-9747AD Groningen, the Netherlands}
\email{estarkenburg@astro.rug.nl}

\author[0000-0002-9086-6398]{Connor Stone}
\affiliation{Mila -- Qu{\'e}bec Artificial Intelligence Institute Ciela Institute Universit{\'e} de Montr{\'e}al}
\email{connorstone628@gmail.com}

\author[0000-0002-0385-0014]{Christopher Storfer}
\affiliation{Institute for Astronomy, University of Hawaii, Honolulu, HI 96822-1897}
\email{cstorfer@hawaii.edu}

\author[0009-0009-9769-534X]{Yoshihisa Suzuki}
\affiliation{Astronomical Institute, Tohoku University, Aoba-ku, Sendai 980-8578, Japan}
\email{yoshihisa.suzuki@astr.tohoku.ac.jp}

\author{Erben, T.}
\affiliation{Argelander-Institut f\"ur Astronomie, Auf dem H\"ugel 71, 53121 Bonn, Germany}
\email{terben@astro.uni-bonn.de}

\author[0000-0001-6469-8805]{Salvatore Taibi}
\affiliation{Institute of Physics, Laboratory of Astrophysics, Ecole Polytechnique F\'ed\'erale de Lausanne (EPFL), 1290 Sauverny, Switzerland}
\email{salvatore.taibi@epfl.ch}

\author[0000-0002-2468-5521]{G. F. Thomas}
\affiliation{Instituto de Astrof\'isica de Canarias, E-38205 La Laguna, Tenerife, Spain}\affiliation{Universidad de La Laguna, Dpto. Astrofísica, E-38206 La Laguna, Tenerife, Spain}
\email{guillaume.thomas.astro@gmail.com}

\author[0000-0002-7032-9667]{Zhang TianFang}
\affiliation{National Astronomical Observatory of Japan, 2-21-1 Osawa, Mitaka, Tokyo 181-8588, Japan}
\email{tianfang.zhang@nao.ac.jp}

\author[0000-0002-3531-7863]{Yoshiki Toba}
\affiliation{Department of Physical Sciences, Ritsumeikan University, Kusatsu, Shiga 525-8577, Japan}\affiliation{National Astronomical Observatory of Japan, 2-21-1 Osawa, Mitaka, Tokyo 181-8588, Japan}\affiliation{Academia Sinica Institute of Astronomy and Astrophysics, 11F of Astronomy-Mathematics Building, AS/NTU, No.1, Section 4, Roosevelt Road, Taipei 10617, Taiwan}
\email{toba@fc.ritsumei.ac.jp}

\author[0000-0002-0673-0632]{Hisakazu Uchiyama}
\affiliation{National Astronomical Observatory of Japan, 2-21-1 Osawa, Mitaka, Tokyo 181-8588, Japan}
\email{huchiyama0326@gmail.com}

\author{David Valls-Gabaud}
\affiliation{LUX, CNRS UMR 8262, Observatoire de Paris, PSL, 61 Avenue de l'Observatoire, 75014 Paris, France}
\email{david.valls-gabaud@obspm.fr}

\author[0000-0003-4134-2042]{Kim A. Venn}
\affiliation{Department of Physics \& Astronomy, University of Victoria, Finnerty Road, Victoria, BC V8P 1A1, Canada}
\email{kvenn@uvic.ca}

\author[0000-0002-2637-8728]{Ludovic Van Waerbeke}
\affiliation{Department of Physics and Astronomy, The University of British Columbia, 6224 Agricultural Road V6T 1Z1, Vancouver, Canada}
\email{waerbeke@phas.ubc.ca}

\author[0000-0002-1341-0952]{Richard J. Wainscoat}
\affiliation{Institute for Astronomy, University of Hawaii, 2680 Woodlawn Drive, Honolulu, HI 96822, USA}
\email{rjw@hawaii.edu}

\author[0000-0002-3303-4077]{Scott Wilkinson}
\affiliation{Department of Physics \& Astronomy, University of Victoria, Finnerty Road, Victoria, BC V8P 1A1, Canada}
\email{swilkinson@uvic.ca}

\author[0000-0002-8173-3438]{Anna Wittje}
\affiliation{Ruhr University Bochum, Faculty of Physics and Astronomy, Astronomical Institute (AIRUB), German Centre for Cosmological Lensing (GCCL), 44780 Bochum, Germany}
\email{awitt@astro.ruhr-uni-bochum.de}

\author{Taketo Yoshida}
\affiliation{Graduate School of Science and Engineering, Ehime University, 3 Bunkyo-cho, Matsuyama-shi, Ehime, Japan}
\email{yoshida@cosmos.phys.sci.ehime-u.ac.jp}

\author[0009-0001-3910-2288]{Yuxing Zhong}
\affiliation{Faculty of Science and Engineering, Waseda University, 3-4-1, Okubo, Shinjuku, Tokyo 169-8555, Japan}
\email{yuxing.zhong.astro@gmail.com}



\begin{abstract}

The Ultraviolet Near-Infrared Optical Northern Survey (UNIONS) is a “collaboration of collaborations” that is using the Canada-France-Hawai'i Telescope, the Pan-STARRS telescopes, and the Subaru Observatory to obtain $ugriz$ images of a core survey region of  6250 deg$^2$ of the northern sky. The $10\sigma$ point source depth of the data, as measured within a 2-arcsecond diameter aperture, are $[u,g,r,i,z] = [23.7, 24.5, 24.2, 23.8, 23.3]$\ in AB magnitudes. UNIONS is addressing some of the most fundamental questions in astronomy, including the properties of dark matter, the growth of structure in the Universe from the very smallest galaxies to large-scale structure, and the assembly of the Milky Way. It is set to become the major ground-based legacy survey for the northern hemisphere for the next decade and provides an essential northern complement to the static-sky science of the Vera C. Rubin Observatory's Legacy Survey of Space and Time. UNIONS supports the core science mission of the {\it Euclid} space mission by providing the data necessary in the northern hemisphere for the calibration of the wavelength dependence of the {\it Euclid} point-spread function and derivation of photometric redshifts in the North Galactic Cap. This region contains the highest quality sky for {\it Euclid}, with low backgrounds from the zodiacal light, stellar density, extinction, and emission from Galactic cirrus. Here, we describe the UNIONS survey components, science goals, data products, and the current status of the overall program.
\end{abstract}


\section{Introduction} 

The Universe is an all-sky object. On the largest scales, it is likely isotropic and homogeneous, but there are significant scientific shortcomings in mistaking this principle of cosmology for one about all of astronomy. It is a matter of considerable excitement that the coming few years will see the start of the Vera C. Rubin's Legacy Survey of Space and Time \citep[LSST;][]{2019ApJ...873..111I}, and usher in a new era of optical wide field surveys from the ground. Combine this with other wavelengths, {\it e.g.} the Square Kilometer Array \cite[SKA;][]{ska2015}, and it is clear that the southern hemisphere is well-served for multi-wavelength imaging surveys for the coming few decades. The same is not necessarily true of the northern sky, however. This is despite the legacy set by the Sloan Digital Sky Survey \citep[SDSS;][]{2000AJ....120.1579Y} and the Pan-STARRS $3\pi$ survey \citep{2016arXiv161205560C} over the last 20 years. Premier astronomical data obtained from space including {\it Gaia} \citep{gaia2016}, the James Webb Space Telescope \citep[JWST;][]{jwst2023} and {\it Euclid} \citep{euclid2024} - to name just a few, currently active, high profile optical/near-infrared telescopes - are hemispherically agnostic, but this is not the case for the ground-based data with which they are often successfully combined. For the local (and even the low redshift) Universe, there are very obvious differences in scientific targets available in the northern and southern skies, and multi-messengers from the very distant Universe do not care about the orientation of the Earth's rotation axis. These considerations suggest that even partially addressing the impending hemispheric imbalance in deep ground-based optical imaging surveys has the significant possibility of enabling new scientific opportunities over the coming decade.

The Ultraviolet Near-Infrared Optical Northern Survey (UNIONS) is a collaboration of wide field imaging surveys of the northern hemisphere that use three observing facilities located in Hawai'i to obtain deep $ugriz$ imaging over much of the northern extragalactic sky. Specifically, the {\it Canada-France Imaging Survey} (CFIS), using MegaCam at the 3.6-meter Canada-France-Hawaii Telescope (CFHT) on Maunakea, is obtaining deep $u$- and $r$-band imaging; Pan-STARRS is obtaining deep $i$- and $z$-band imaging; Hyper Suprime-Cam mounted on the Subaru 8.2-meter telescope, is obtaining deep $z$-band imaging through the {\it Wide Imaging with Subaru HSC of the Euclid Sky} (WISHES) and deep $g$-band imaging through the {\it Waterloo-Hawai'i Institute for Astronomy $g$-band Survey} (WHIGS). These efforts are directed, in part, to securing optical imaging to complement the Euclid space mission \citep{euclid2024}, although UNIONS is a separate collaboration aimed at maximizing the science return of these large and deep surveys of the northern skies.

At the time of writing, UNIONS data have formed the basis for more than 30 publications currently in the peer-reviewed literature. Further, while the original survey area is near completion, extensions to the survey have recently been approved to push to lower declination. As such, the intent of this current contribution is to describe the survey components, the collaboration, the science goals, the data products, and the current status of the overall program.

Section 2 provides an overview of UNIONS, including the primary science drivers and overall survey characteristics, including a brief summary of relevant details necessary to understand the survey footprint. Section 3 provides the survey strategy and data reduction details of each of the main survey components of UNIONS. Section 4 describes the methodology and characteristics of the overall multi-band catalog that is being created from UNIONS data. Section 5 briefly describes the {\it Euclid} data products that will be derived from UNIONS data and differentiates between them and the UNIONS data products, and Section 6 provides a summary.

\section{Overview of UNIONS}
\label{sec:overview}

UNIONS has emerged over the last several years, as various teams recognized the powerful scientific and legacy opportunities of uniting their separate efforts into a joint program of activities. In Section~\ref{ssec:science}, we describe some of the driving science of UNIONS. In Section~\ref{ssec:character}, we summarize the defining characteristics of the UNIONS survey, including a few necessary details on its formation in order to understand the definition of the survey footprint. 

\subsection{UNIONS science}
\label{ssec:science}

UNIONS is a community legacy survey that provides a strong scientific platform for a wealth of astrophysics. It is helping to answer some of the most fundamental questions in astronomy, including the properties of dark matter, the growth of structure in the Universe from Galactic to cluster scales, the assembly of the Milky Way, and it will culminate by directly contributing to the state-of-the-art dark energy measurement expected from the {\it Euclid} mission. In the new era of time domain and multi-messenger astrophysics, it is more important than ever to build up a high-quality, multi-band archive of the sky to ensure a record exists of the progenitor systems that spawn the most exciting astrophysical transients. Here, we summarize a few key science themes of UNIONS and highlight some published results that stem from UNIONS data taken to date. We note that, in very broad terms, UNIONS data are only slightly shallower in photometric depth than Year 1 of the LSST. This enables an extensive number of static sky science cases.

\subsubsection{Galactic Archaeology} 

A key original science driver of the survey that became UNIONS is near field cosmology. The $u$-band data of UNIONS are approximately 3 magnitudes deeper than for SDSS and are designed to take advantage of the blue sensitivity of CFHT compared to much larger optical ground-based telescopes. The $u$ band, in combination with other optical bands, allows for highly accurate estimates of photometric distances \citep{ivezic2008} that have significantly higher precision than {\it Gaia} astrometry at large ($>$\.10\.kpc) distances, making this an ideal approach for pushing {\it Gaia} proper motions into the outer stellar halo, where {\it Gaia} parallaxes are poor. Prior to obtaining equivalently deep bands with the rest of UNIONS, early science in this area involved combining the $u$-band data with other deep photometric and astrometric surveys, especially Pan-STARRS $3\pi$ and {\it Gaia}. The first results were presented in \cite{ibata2017a, ibata2017b}, and an enhanced methodology for stellar distance determination using photometry alone is presented in \cite{thomas2019b}. With these data, we have been able to discover different accreted structures, such as streams around M92 \citep{thomas2020} and NGC\,5466 \citep{jensen2021}, and we have probed the structure of our Galaxy at different scales from the Galactic disk \citep{thomas2019b,2025arXiv250217319B} to the outskirts of the stellar halo of the Milky Way \citep{thomas2018}. We have also explored the white dwarf population of the Milky Way (\citealt{fantin2021}), including providing a novel method for determining the star-formation history of the disk (\citealt{fantin2019}).

The acquisition of equivalently deep $gri$ data to complement the $u$ band as part of UNIONS has enabled new searches for Milky Way and Local Group satellites. Here, we take advantage of the deep, wide-field nature of UNIONS to search for resolved overdensities of stars (i.e., dwarf galaxies and star clusters) in the Milky Way halo, M31 halo, and Local Group. Three discoveries have been published: B{\"o}otes V (\citealt{smith2023}; see also \citealt{cerny2023}) is an ultra-faint, very metal-poor, dwarf galaxy; Ursa Major III/UNIONS 1 \citep{smith2024} is the faintest known satellite of the Milky Way with a stellar mass of only about $16\,M_\odot$, whose stellar velocities may indicate the presence of dark matter, though additional work is being done to better understand this remarkable object \citep{errani2024, crnogorvevic2024}; and Pegasus VII (\citealt{smith2025}) is a new satellite of the M31 galaxy. Studies of three additional Milky Way satellites are currently being prepared for publication.

The combination of UNIONS $u$-band data with narrow-band Ca H\&K photometry from the CFHT Pristine survey \citep{Starkenburg2017, Martin2024} enables even more Galactic Archaeology science. For example, the combination of both bands provide a 3D view of the outermost Galactic halo through a clean selection of blue horizontal branch stars \citep{titulaer2021} that have been used to study the outermost spur feature of the Sagittarius dwarf galaxy stellar stream at $\sim$140\,kpc and provide additional constraints on the Galactic gravitational potential at those distances \citep{Bayer2025}. 

\subsubsection{Weak Gravitational Lensing} 
Another major scientific objective of UNIONS is weak gravitational lensing, which enables the study of dark matter distribution across a wide range of scales, from individual galaxies to large-scale cosmic structures. Additionally, by cross-correlating with baryonic probes, it allows for the exploration of the relationship between dark matter and baryonic matter.

Prior to the advent of the full multi-band version of UNIONS, the very high image quality of the CFHT $r$ band was a major factor in the development of the science case that enabled these observations. It was recognized that the completed $r-$band data has key advantages over competing weak lensing surveys, such as the Dark Energy Survey \citep[DES;][]{GattiSheldonAmon2021}. The areas are similar (about 5000~deg$^2$), but the significantly better seeing at CFHT (the median $r-$band image quality is 0.7$^{\prime\prime}$) allows us to make shape measurements of smaller galaxies, providing a higher background source density, and so improving the lensing power statistics.

For UNIONS, we have developed a weak lensing shape measurement pipeline, \textsc{ShapePipe} \citep{GuinotKilbingerFarrens2022, FarrensGuinotKilbinger2022}, which is based on \textsc{ngmix} \citep{Sheldon2015} and \textsc{Metacalibration} \citep{HuffMandelbaum2017, SheldonHuff2017}. New methods to characterize the point-spread function (PSF) have also been developed \citep{LiaudatBonninStarck2021} as well as novel methods to characterize the PSF \citep{2024arXiv241214666G,ZhangKilbingerHervasPeters2024}. The lensing catalog that is currently used internally (not based on the full data set) spans 3500~deg$^2$ and has an effective background source density of 7 to 8 galaxies per arcmin$^2$.  For comparison, DES-Year-3 has an effective background source density of 5.6 galaxies per arcmin$^2$ \citep{GattiSheldonAmon2021}. An upcoming UNIONS shape catalog that covers a larger area, with a slightly higher source density as well as photometric redshifts for the sources, is under active development.

Another critical advantage for UNIONS weak lensing is the vast wealth of spectroscopy in the UNIONS foreground due to the SDSS surveys and, most recently, the Dark Energy Spectroscopic Instrument survey \citep[DESI,][]{desi2016}. This opens up a huge range of science currently impossible in the southern hemisphere. An important early result is the first measurement of the black hole-halo mass relation \citep{LiKilbingerLuo2024, 2024ApJ...969L..25L} based on weak lensing. We have also studied the ellipticity of dark matter halos around luminous red galaxies from SDSS \citep{RobisonHudsonCuillandre2023}. 

In cosmology, we are using the spectroscopic samples from the  Baryon Oscillation Spectroscopic Survey \citep[BOSS,][]{2013AJ....145...10D} and DESI to construct $3 \times 2$-point statistics (cosmic shear, galaxy-galaxy lensing, and galaxy clustering) to obtain tight constraints on the cosmological parameters. We intend to further tighten these using higher order methods such as density-split statistics \citep{BurgerPaillasHudson2024} or weak lensing peak counts \citep{AycoberryAjaniGuinot2023}. Other projects recently completed include measuring the weak (anti-)lensing signature from spectroscopically-identified cosmic voids and a direct measurement of ``intrinsic alignments'' -- the tendency of galaxies to point towards each other -- from BOSS spectra \citep{2024arXiv241201790H}.  Cosmic shear, which examines dark matter statistics on the largest scales, is another powerful cosmological probe made possible by the unprecedented area and depth of UNIONS. Finally, clusters of galaxies and their splashback radius are another potential cosmological probe as discussed in \citet{2025arXiv250109147M}.

Finally, weak lensing can also be used to probe the link between dark matter and galaxy formation and evolution. We are studying the stellar-mass-halo-mass relation as a function of galaxy color and redshift and testing for dependence on ``third parameters'' such as galaxy size or large scale environmental overdensity (``assembly bias'').  We also intend to study halos of unusual galaxies, such as ultra-diffuse dwarfs. Lastly, we can study more complicated structures, such as filaments of the cosmic web between spectroscopically-identified galaxy pairs \citep[following, e.g.][]{EppsHudson2017}, both in terms of their dark matter structure \citep{YangHudsonAfshordi2022} and in cross-correlation with baryons probed by the thermal and kinetic Sunyaev-Zel’dovich effect. \citep{2022A&A...660A..27T,2023MNRAS.520..583J,2020ApJ...890..148U,2017MNRAS.467.2706T,2024MNRAS.534..655B}

\subsubsection{Legacy imaging for the northern hemisphere} 

The scope of UNIONS is such that it is becoming the major wide-field optical survey of the northern sky for the 2020s and into the era of the Vera C. Rubin Observatory. Even when the LSST is fully underway, UNIONS will remain unsurpassed from the ground in the northern hemisphere. Critically, UNIONS is helping to address the hemispheric imbalance in wide field imaging that will otherwise limit the utility of ground-based optical surveys to complement other multi-wavelength, multi-messenger, and space-based missions. In particular we highlight the following three motivations for UNIONS:

\begin{itemize}

\item 
Deep multi-band imaging in the north is an essential complement to
multi-wavelength astronomy enterprise. It enables the identification
of photometric sources for combination with past and ongoing
ground-based wide-area surveys in the north, such as the LOFAR
Two-meter Sky Survey (LoTSS) at 144 MHz \citep{2022A&A...659A...1S},
VLA/FIRST and NVSS at 1.4 GHz
\citep{2015ApJ...801...26H, 1998AJ....115.1693C}, and VLASS at 3 GHz
\citep{2021ApJS..255...30G}. Future surveys will
further expand northern coverage in more deep, including those by the
Next Generation Very Large Array
\citep[ngVLA;][]{ngvla2018}
and the
Square Kilometre Array (SKA), which is not restricted to southern
declinations.
Additionally, space-based all-sky missions, including
WISE \citep{2010AJ....140.1868W} in the mid-infrared, in the mid- and far-infrared, and eROSITA
\citep{2021A&A...647A...1P, 2024A&A...682A..34M}
at X-ray wavelengths are
hemispherically agnostic, reinforcing the importance of completing
UNIONS in the north to provide the necessary ground-based optical
context.

\item In the era of multi-messenger astronomy, gravitational wave detections can come from anywhere on the sky. It is imperative that we have deep imaging already available to provide the necessary difference imaging to close in on the nature of the astrophysical sources (for example, see \citealt{smartt2017}).

\item The attention of wide field optical astronomy is turning towards spectroscopy: the Dark Energy Spectroscopic Instrument \citep[DESI;][]{desi2016} is delivering a groundbreaking spectroscopic data set in the north \citep{desi_results_I_2024, desi_results_II_2024, desi_results_III_2024}, soon to be followed by the WHT Enhanced Area Velocity Explorer \citep[WEAVE;][]{weave2012} and the Prime Focus Spectrograph on the Subaru Telescope\citep{2022SPIE12184E..6RW} in the north, and 4MOST \citep{4most2019} and VLT/MOONS \citep{moons2020} in the south. UNIONS provides an extremely robust foundation for large aperture northern
hemisphere wide field spectroscopy by providing a high quality source of multi-band pre-imaging that otherwise does not exist over most of the northern sky. For example the $u$-band photometry is necessary for target selection of Lyman break galaxies (LBGs), a key high-z tracer in DESI-II and other spectroscopic surveys such as spec-s5, MSE and MUST \citep{2024arXiv241008062P}.
\end{itemize}

The broad science goals of such a data set is now being reflected in the publications that have emerged based on UNIONS data. In addition to the topics discussed in earlier sections, UNIONS data have been used to study the role of ram pressure stripping in galaxies (\citealt{2025arXiv250213123F,roberts2022}), to examine the role and impact of mergers on galaxy evolution (\citealt{ellison2019, ellison2022,ellison2024,bickley2021, bickley2022, bickley2023, wilkinson2022,  ferreira2024, ferreira2025}), to understand processes leading to the formation of fossil groups (\citealt{chu2023}), to characterize diffuse low surface brightness structures around galaxies (\citealt{sola2022}), to probe the cosmic far infrared background (\citealt{lim2023}), and to search for strong lensing events \citep{savary2022,2025arXiv250310610A}, including distant quasars (\citealt{chan2022}).

\subsubsection{Connection with {\it Euclid}}
\label{sssec:unionseuclid}

In addition to stand-alone science, UNIONS provides color information to the {\it Euclid} survey.  Briefly, {\it Euclid} is obtaining deep imaging of 13~000~deg$^2$ of sky in one broad optical band (VIS) and three (YJH) near-infrared bands, as well as grism spectroscopy in the near-infrared. Its core science mission is based around weak lensing and galaxy clustering. Measuring the weak lensing signal with the necessary degree of precision requires detailed knowledge of the PSF in the VIS band. Since the PSF is wavelength dependent, modelling the PSF at the level needed requires knowledge of the spectral energy distribution of the sources. This in turn requires multi-band photometry over the spectral range of the VIS bandpass (see \citet{2018MNRAS.477.3433E} for more details) to model the wavelength dependence of the PSF. This then impacts measurements of the galaxies for weak lensing, where the model PSF needs to be applied in a wavelength-dependent way to accurately measure the spatial profile of the galaxy. In addition, UNIONS photometry is combined with the {\it Euclid} YJH photometry to detemine photometric redshifts which allow cosmological tomography, turning the monochrome, two-dimensional {\it Euclid} VIS images into a multi-color, three dimensional picture of the universe. To achieve its goals, {\it Euclid} requires photometric redshift errors of $\sigma_z/(1+z)<0.05$.

In the south, color information for {\it Euclid} is presently provided by the Kilo-Degree Survey \citep[KiDS]{2017A&A...604A.134D} and the Dark Energy Survey \citep[DES]{2018ApJS..239...18A}. These surveys will eventually be superseded by the LSST \citep{2019ApJ...873..111I} while north of a Declination of 15 degreees, the $ugriz$ data needed by {\it Euclid} is provided by UNIONS.

\subsection{Survey characteristics}
\label{ssec:character}

\subsubsection{Survey footprint}
\label{sec:footprint}

\begin{figure*}
  \begin{center}+
      \includegraphics[width=0.95\hsize]{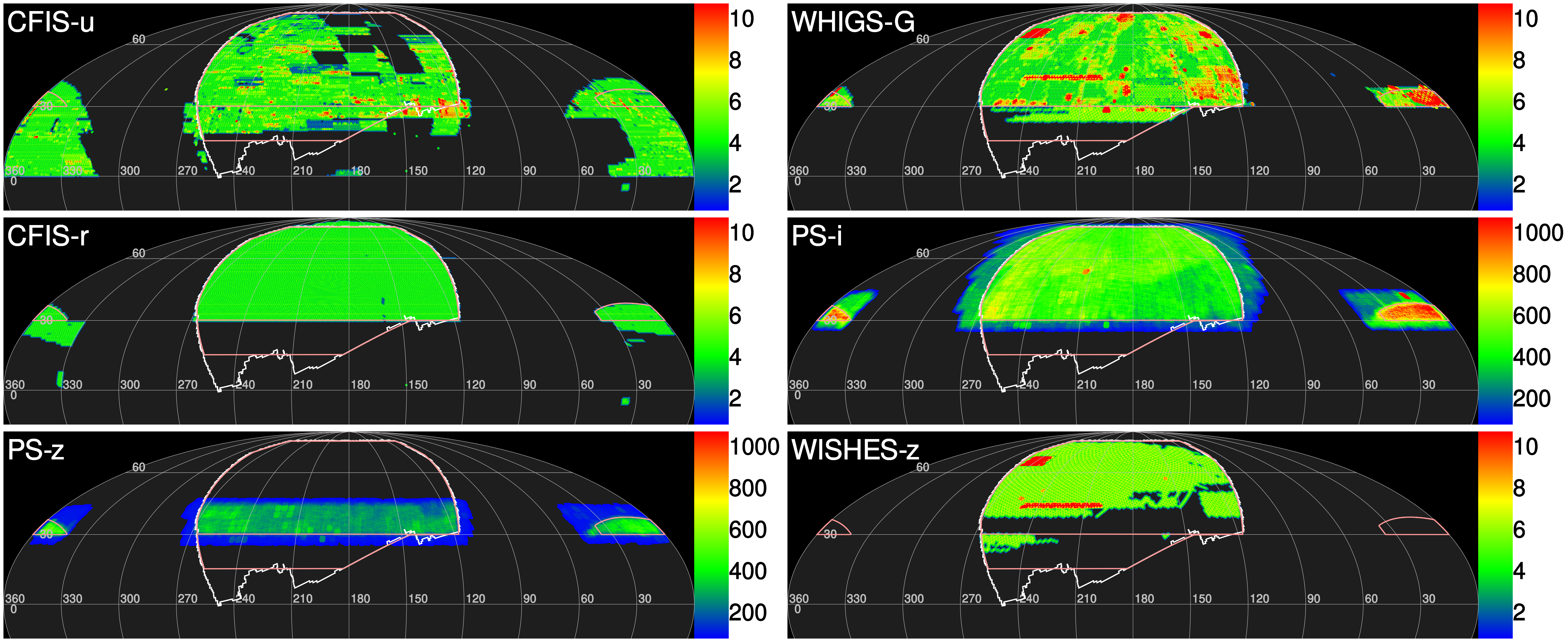}
  \end{center}
  \caption{Current spatial coverage of UNIONS for each of the five bands. The color scale indicates the number of images acquired (as of January 2025) in each location. Green (through red) indicates that the nominal depth has been achieved. The red lines show the nominal UNIONS footprint, both the current target footprint (down to $\delta = 30^\circ$) and the proposed extension (down to $\delta = 15^\circ$). The white line indicates the full northern {\it Euclid} Wide Survey footprint. The figure uses a Mollweide equal-area projection.}
\label{fig:coverage}
\end{figure*}

Figure~\ref{fig:coverage} shows the spatial coverage as of January 2025 for each of the five bands of UNIONS. The survey is primarily focused on the northern sky away from the Galactic plane, $|b| \ge 25^\circ$. The red lines show the primary UNIONS survey region. The {\it Euclid} final data release (DR6) region  that covers the North Galactic Cap (NGC) is shown by a white line. In this figure, the color-scale shows the number of individual exposures contributing to each area of sky. 

The footprint of UNIONS reflects an evolution in the science goals and scope of the component surveys. What became UNIONS originally started out as a CFHT/MegaCam $u-$band survey called the {\it Legacy for the U-band All-sky Universe}, LUAU \citep{2017ApJ...848..128I}, primarily focused towards Galactic archaeology. At around the same time, the {\it Euclid} community started  looking for options to obtain northern ground-based imaging necessary for their core science.  The rationale for the specific {\it Euclid} survey area is given in detail in \cite{euclid2024}. As a result of this {\it Euclid}-driven effort, LUAU was subsumed into a new CFHT program, the Canada-France Imaging Survey (CFIS), continuing the $u$~band and also now incorporating $r$~band coverage over the {\it Euclid} northern footprint.

The success of CFIS led to another successful {\it Euclid}-related program using the Pan-STARRS telescopes for the $i$ and $z$ bands. Subsequently, the WISHES team was successful in its application to use Subaru/HSC to obtain $z$-band data for {\it Euclid}. This helped offset some of the Pan-STARRS $z$-band effort, since the observing efficiency for Pan-STARRS in this band is not particularly high. Given the obvious complementarity of the surveys, the three observing programs united under the UNIONS banner, enabling both {\it Euclid} and stand-alone science programs. Given the absence of $g$-band, the UNIONS collaboration pursued successful observing time on Subaru/HSC via the Institute for Astronomy in Hawai'i and Canadian Gemini exchange time, leading to the WHIGS program. 

We note that the {\it Euclid}-relevant area of UNIONS has changed during the lifetime of these observing programs. When CFIS began, the {\it Euclid} Wide Survey footprint originally included the region around the southern Galactic cap, which is why this area is also covered in all five UNIONS bands (although not by the WISHES $z$ band). Further, it was originally thought that Rubin/LSST would likely provide data for {\it Euclid} to the necessary depth up to  $\delta = 30^\circ$, and so primary attention was given to the region north of this. However, in recent years, it has become clear that Rubin/LSST will only extend to $\delta = 15^\circ$. As such, the UNIONS programs have sought extensions to obtain complete coverage of the region outlined in red in Figure~\ref{fig:coverage} that extends below  $\delta = 30^\circ$. These extension programs have all been granted and begin in 2025. The extended area is shown by the second red line on Figure \ref{fig:coverage}.

With reference to Figure~1, the {\it Euclid} survey region around the North Galactic Cap at $\delta \ge 30^\circ$ is 4606~deg$^2$. The {\it Euclid} survey region around the North Galactic Cap in the range $15 \le \delta \le 30^\circ$ is 1381~deg$^2$. The region around the South Galactic Cap (SGC) originally (but no longer) included in {\it Euclid}, and observed by UNIONS, is 273~deg$^2$. These three regions make up the core UNIONS survey area. Table~\ref{tab:areas} lists the fraction of these three areas surveyed by UNIONS as of January 2025. Also included in the final column is additional area that has been observed to comparable depth and which is included in the final data set, and which is visible in the figure. We note that full coverage of the {\it Euclid} area in the $z$-band is provided by the combination of WISHES and Pan-STARRS data, with the overlap region between these programs occurring at $38^\circ \lesssim \delta \lesssim 42^\circ$.

\begin{table*}
\label{tab:areas}
\begin{center}
\caption{Summary of the coverage, as of January 2025, of the component surveys contributing to UNIONS, given in deg$2$. The columns show the coverage in the two North Galactic Cap (NGC) areas and the South Galactic Cap (SGC) area as measured in deg$^2$ and as a percentage completeness of each area. The Extra column gives any additional survey area in deg$^2$ outside of these regions. Full $z$-band coverage is obtained by combining both the Pan-STARRS and WISHES data sets. Note that Pan-STARRS $i$-band coverage is complete in terms of area (north of $\delta=30^\circ$) but not depth.}
\begin{tabular*}{\textwidth}{llcccccc}
\hline
Filter & Survey & Instrument &
NGC, $\delta \ge 30^\circ$ & 
NGC, $\delta < 30^\circ$ & 
SGC  & Extra & Total \\
\hline
      $u$ &            CFIS &    CFHT/MegaCam &    3732 (81\%)  &    1089 (78\%) &    273 (100\%) &   3233 &   8327 \\ 
       $g$ &           WHIGS &      Subaru/HSC &    4585 (99\%)  &     627 (45\%) &    273 (100\%) &    270 &   5755 \\ 
       $r$ &            CFIS &    CFHT/MegaCam &    4582 (99\%)  &      91 (6\%) &    273 (100\%) &    692 &   5638 \\ 
       $i$ &      Pan-STARRS &      Pan-STARRS &    4606 (85\%)  &     513 (37\%) &    273 (100\%) &   2325 &   7717 \\ 
  $z_{\rm PS}$ &  Pan-STARRS &      Pan-STARRS &    2070 (44\%)  &     502 (36\%) &    273 (100\%) &   1143 &   3988 \\ 
 $z_{\rm HSC}$ &      WISHES &      Subaru/HSC &    3266 (70\%)  &     216 (15\%) &      0 (0\%) &     32 &   3514 \\ 
\hline
\end{tabular*}

\end{center} 
\end{table*}

\subsubsection{Filters, depth and image quality}
\label{sec:filters}

\begin{figure}
  \begin{center}
      \includegraphics[width=0.95\hsize]{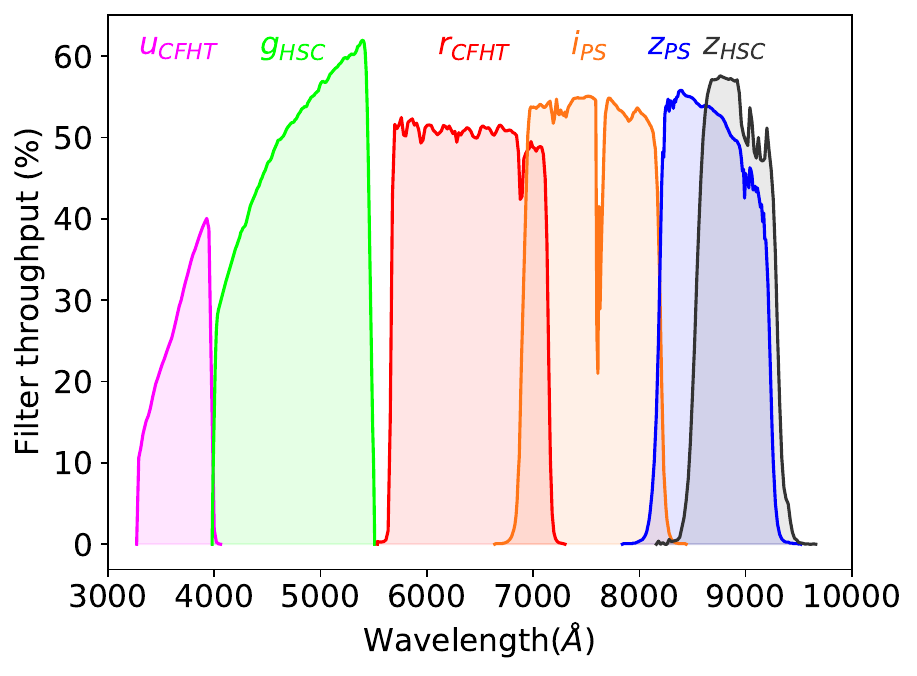}
  \end{center}
  \caption{Transmission curves of the filters used by UNIONS. These correspond to the effective bandpass accounting for the filter transmission as well as the full system throughput.}
\label{fig:transmit}
\end{figure}

The filter transmission functions for each of the survey components of UNIONS are shown in Figure~\ref{fig:transmit}. The central wavelengths, blue-end (“cut-on”) and red-end (“cut-off”) wavelengths are listed in Table~\ref{tab:filters}. The transmission functions are available online in text format\footnote{\url{https://www.canfar.net/storage/vault/list/cfis/filters}}, and in each case contain the full system response. Specifically, for CFHT, the MegaCam filters introduced in 2014 are used, u.MP9302 and r.MP9602\footnote{\url{https://www.cfht.hawaii.edu/Instruments/Filters/megaprime.html}}. Their response curves include the MegaCam quantum efficiency, the transmission of the MegaPrime camera optics, the reflectivity of the CFHT primary mirror and a nominal atmosphere of 1.2 airmasses\footnote{\url{https://www.cadc-ccda.hia-iha.nrc-cnrc.gc.ca/en/megapipe/docs/filt.html}}. The Pan-STARRS filter set, together with the responses of the telescope, camera and detectors, are described in considerable detail in \citet{2012ApJ...750...99T}. Similarly, the HSC filters and the response functions of the HSC system are described in \citet{2018PASJ...70...66K}. UNIONS magnitudes are computed on the AB system \citep{1974ApJS...27...21O}.

\begin{table*}
    \centering
             \caption{UNIONS filter properties. The table gives the filter name matched with the survey, the filter name as used by the respective telescope and the center cut-on, and cut-off values}
  \begin{tabular}{ccccccc}
   		\hline
     Survey filter name & Filter name &Centre (\AA) & 50\% cut-on (\AA) & 50\% cut-off (\AA) \\
		\hline
    CFIS $u$       & u.MP9302      &     3682 &      3470 &     4000 \\ 
    WHIGS $g$      & HSC-g         &     4816 &      4090 &     5470 \\ 
    CFIS $r$       & r.MP9602      &     6425 &      5677 &     7146 \\ 
    Pan-STARRS $i$ & $i_{\rm PS}$  &     7544 &      6910 &     8200 \\ 
    Pan-STARRS $z$ & $z_{\rm PS}$  &     8679 &      8190 &     9220 \\ 
    WISHES $z$     & HSC-z         &     8903 &      8540 &     9300 \\ 
		\hline\\
    \end{tabular}
 	\label{tab:filters}
\end{table*}

\begin{figure*}
  \begin{center}
      \includegraphics[width=0.95\hsize]{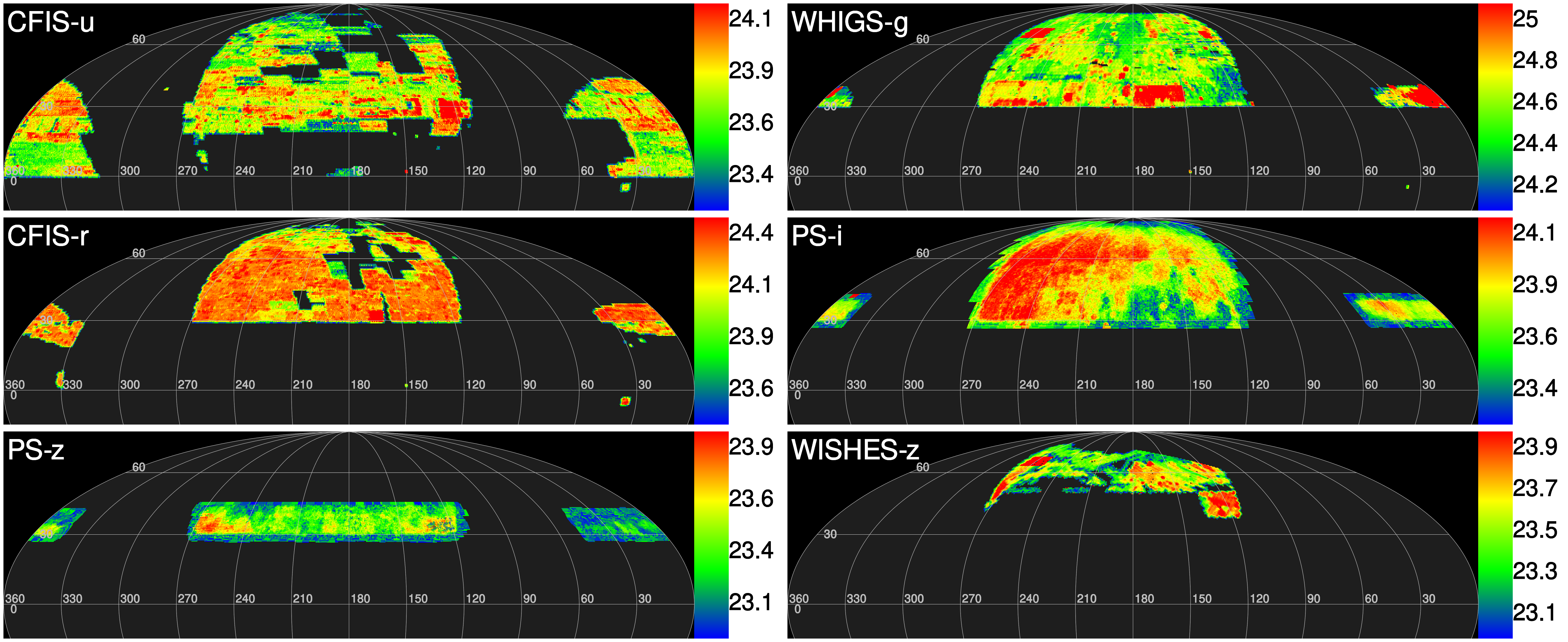}
  \end{center}
  \caption{Map of the current measured depth for the UNIONS survey, quantified as the 10$\sigma$ point source depth measured through a $2^{\prime\prime}$ aperture. Note the color scale varies for each filter, where the median depth is shown as green.}
\label{fig:2Ddepths}
\end{figure*}

\begin{figure*}
  \begin{center}
      \includegraphics[width=8.5cm]{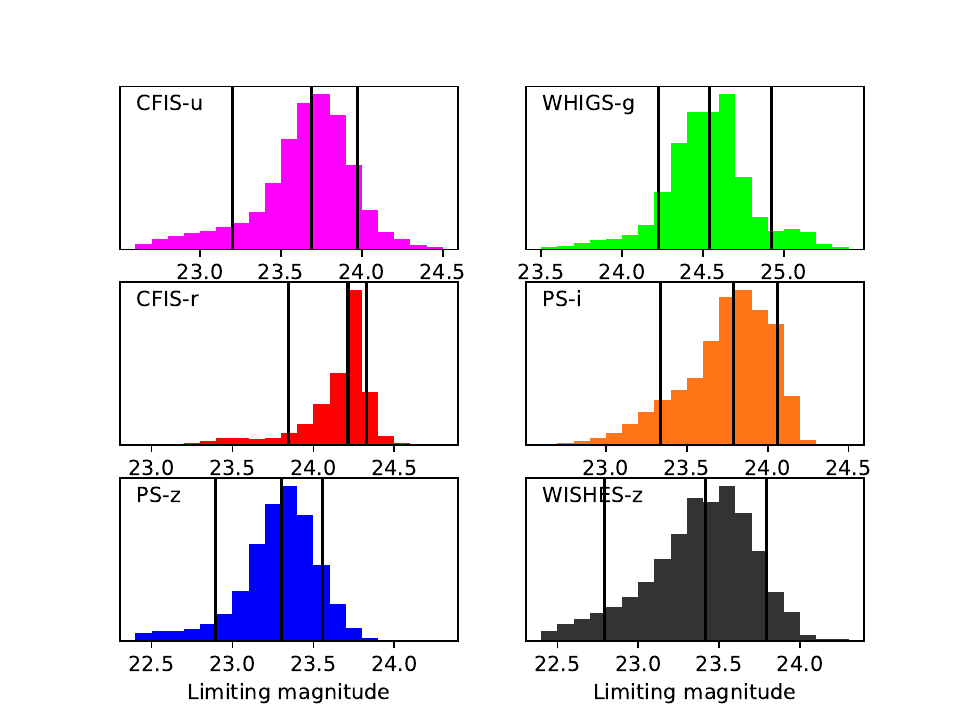}
      \includegraphics[width=8.5cm]{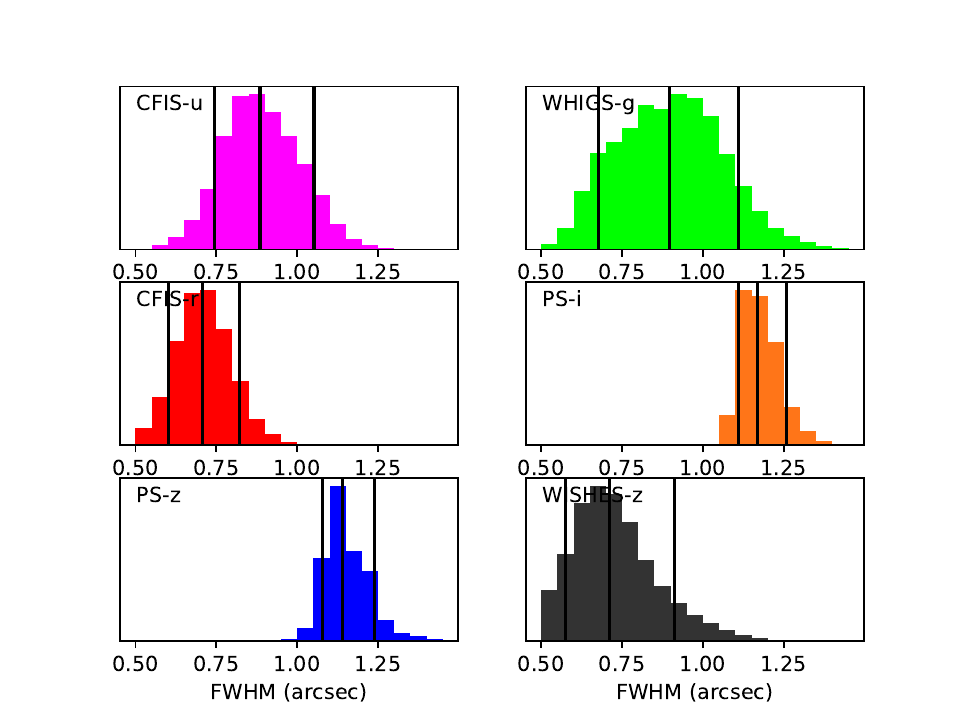}
  \end{center}
  \caption{Left panels: Histograms of the 10$\sigma$ point source depths of each UNIONS band measured through a 2$^{\prime\prime}$ aperture.  Right panels: Histograms of the image quality for each UNIONS band, measured as the FWHM of point sources. In all panels, the vertical lines show the median value and the 10th and 90th percentiles. }
\label{fig:histograms}
\end{figure*}

We characterize the depth of UNIONS photometry as the $10\,\sigma$ point source depth measured through a $2^{\prime\prime}$ aperture. This particular choice of depth metric is driven by specifications from the {\it Euclid} survey. For telescopes which enjoy the superb seeing conditions of Maunakea, it is somewhat conservative. Depths computed through optimal apertures will be systematically deeper than the ones reported here. The depths are calculated as a function of position and are shown for each filter in Figure~\ref{fig:2Ddepths}. Histograms of these values are shown for each filter in the left panels of Figure~\ref{fig:histograms}, with vertical lines indicating the median depth and the 10th and 90th percentiles values. These values are given explicitly in Table~\ref{tab:ml}. The current median depth of UNIONS is $[u, g, r, i, z_{\rm PS}, z_{\rm HSC}] = [23.7, 24.5, 24.2, 23.8, 23.3], 23.4$, to be compared with the original goal of $[u, g, r, i, z] = [23.6, 24.5, 23.9, 23.6, 23.4]$. We note that the depth of the $ugrz_{\rm PS}z_{\rm HSC}$ data is unlikely to change notably as new area is added. However, the survey strategy of Pan-STARRS is such that we can expect the depth and uniformity of the $i$-band data to improve (although no future $z_{\rm PS}$ data is expected to be added).

\begin{table}
     \caption{10-sigma point source depths of UNIONS measured through a 2-arcsecond aperture.}
    \begin{center}
  \begin{tabular}{cccc}
   		\hline
     Filter name & 10\%-ile & Median& 90\%-ile \\
		\hline
         CFIS $u$  &  23.20 &   23.69 &   23.97 \\
        WHIGS $g$  &  24.22 &   24.54 &   24.92 \\
         CFIS $r$  &  23.84 &   24.21 &   24.33 \\
   Pan-STARRS $i$  &  23.34 &   23.79 &   24.06 \\
   Pan-STARRS $z$  &  22.89 &   23.31 &   23.56 \\
       WISHES $z$  &  22.79 &   23.41 &   23.80 \\
 		\hline
    \label{tab:ml}
    \end{tabular}
    \end{center}
\end{table}

\begin{figure*}
  \begin{center}
      \includegraphics[width=0.95\hsize]{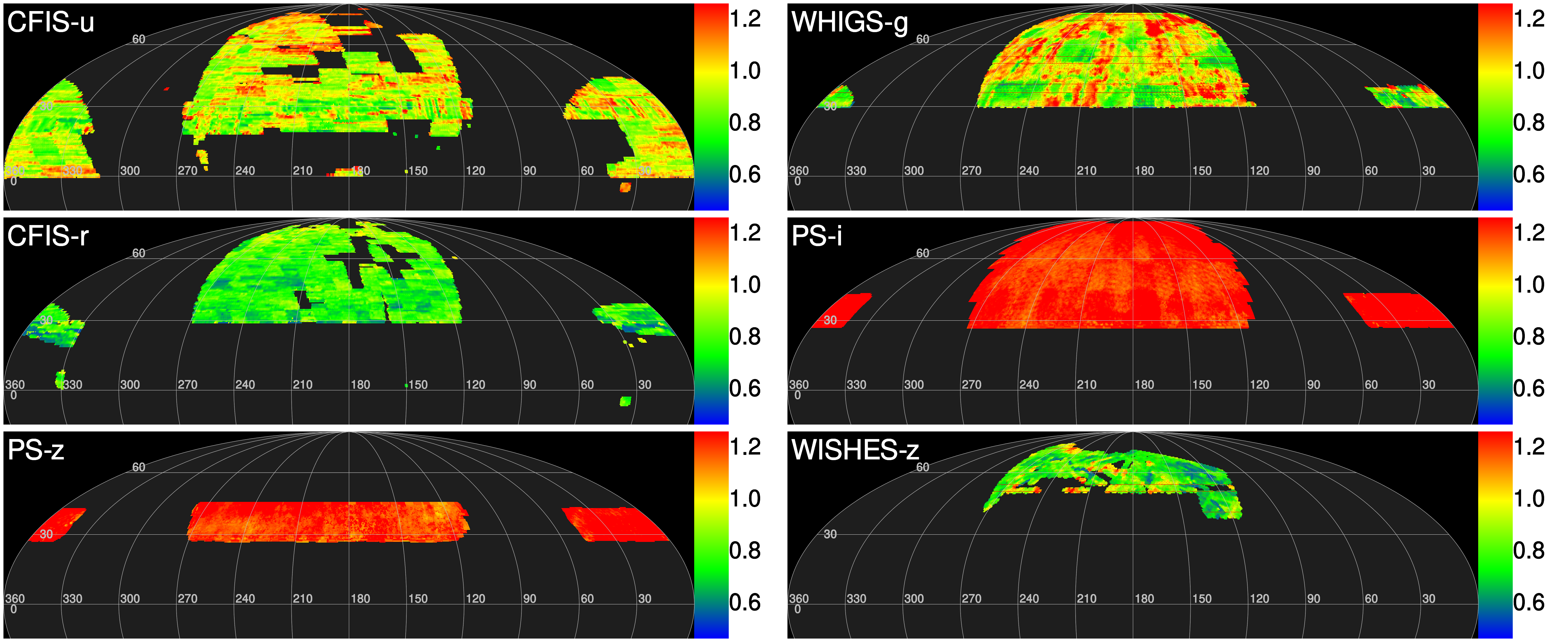}
  \end{center}
  \caption{Maps of the image quality for each UNIONS band, measured as the FWHM of point sources measured in arcseonds across the survey regions.}
\label{fig:2DIQ}
\end{figure*}

Figure~\ref{fig:2DIQ} shows maps of the effective image quality (IQ, specifically, the full width at half maximum of point sources measured in arcseconds) as a function of position for each UNIONS band. The color scales are the same in each panel. The IQ is shown as one-dimensional histograms in the right panel of Figure~\ref{fig:histograms}, with the median, 10th and 90th percentiles marked as vertical lines. These values are given explicitly in Table~\ref{tab:iq}. The survey strategy of Pan-STARRS ensures that many exposures are taken at any particular position in a wide range of observing conditions, so the average IQ of the survey will be uniform by design. The excellent image quality of the UNIONS data across all bands is clear, and the uniformity of the $r-$band data is particularly striking. The quality of the $u$- and the $r$- band data benefit enormously from the queue scheduling at CFHT, as discussed in section \ref{ssec:CFISstrategy}

\begin{table}
    \centering
     \caption{Image quality (FWHM) for each UNIONS band.}
  \begin{tabular}{cccc}
   		\hline
     Filter name & 10\%ile & Median& 90\%ile \\
		\hline
         CFIS $u$  &   0.74 &    0.89 &    1.05 \\
        WHIGS $g$  &   0.68 &    0.90 &    1.11 \\
         CFIS $r$  &   0.60 &    0.71 &    0.82 \\
   Pan-STARRS $i$  &   1.11 &    1.17 &    1.26 \\
   Pan-STARRS $z$  &   1.08 &    1.14 &    1.24 \\
       WISHES $z$  &   0.58 &    0.71 &    0.91 \\  
   		\hline
    \label{tab:iq}
    \end{tabular}
\end{table}

\section{The components of UNIONS}
\label{sec:surveys}

The nature of UNIONS as a multi-program, multi-telescope, endeavor, necessarily requires each component survey to operate quasi-independently in terms of scheduling and observing strategies. All of the component surveys are at an advanced stage, but all still ongoing at the time of writing in January 2025. Here, we provide the observing details, survey strategy and data processing efforts specific to each component of UNIONS.

\subsection{The Canada-France Imaging Survey}

CFIS is conducted on the MegaCam camera \citep{2003SPIE.4841...72B} installed on CFHT.

\subsubsection{Survey strategy} 
\label{ssec:CFISstrategy}
CFHT's flexible QSO-SNR (Queued Service Observations: Signal to Noise Ratio) system allows the survey to specify the seeing conditions, which effectively improves the image quality of the survey. Within UNIONS, the $r$-band data have the best (given its wavelength) and most uniform image quality, as previously shown in Figure~\ref{fig:2DIQ}. The CFHT QSO only observes in the $r$ band when the seeing is below $1^{\prime\prime}$. The weak lensing science drives this constraint. Given the generally superb seeing conditions on Maunakea, this constraint does not impose a significant reduction in available observing opportunities.

The QSO-SNR system also automatically adapts the exposure time as a function of the atmospheric conditions (seeing and transparency) to reach a specified depth, which also explains the uniformity of the sensitivity of the $r$-band data, demonstrated by Figure~\ref{fig:2Ddepths}. While the average $r$-band exposure time is 145 seconds, the exposure times actually range from 100 to 300 seconds. Given the relatively short $u-$band exposure times, we hold these fixed at 80 seconds. 

For each pointing in each band, three images are taken, dithered by $0.33^{\circ}$ in Dec., and $0.35^{\circ}/\cos(\delta)$ in RA. Adjacent pointings are spaced $0.98^{\circ}$ apart in Dec., and $0.97^{\circ}/\cos(\delta)$ in RA. This strategy minimizes area loss due to the gaps between chips, increases uniformity in depth, and maximizes survey speed. To give flexibility to the QSO observing team, the sky coverage is organized by layers of tessellation (batches of the first dither position together, batches of the second dither position together, etc.), and the three visits of a given sky area can be observed on different nights, or even years. This approach has proven highly efficient in helping the QSO team build flexible queues.

The adopted observing strategy naturally enables the low surface brightness (LSB) observing mode with CFHT/MegaCam in the $r$ band \citep[section 4.2]{2012ApJS..200....4F}. Specifically, the telescope moves at least one full field of view between exposures. Chopping between targets in this manner makes it possible to reconstruct the sky background for each exposure, provided a minimum of seven exposures are taken in each observing block. This then enables some key science, for example in detecting and characterizing faint features around galaxies and merger remnants (e.g., see results in \citealt{sola2022}). 

\begin{figure}
  \begin{center}
      \includegraphics[width=0.95\hsize]{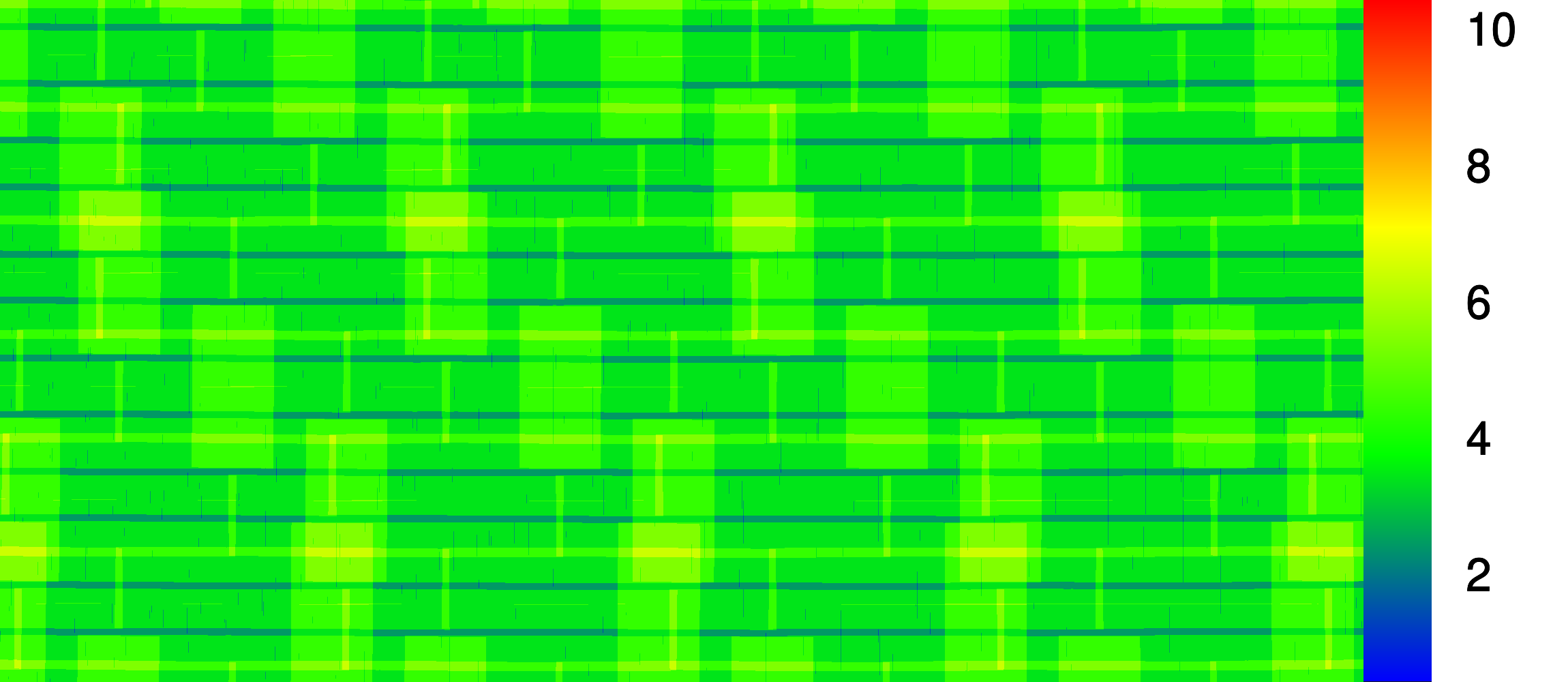}\vspace{0.5cm}
      \includegraphics[width=0.95\hsize]{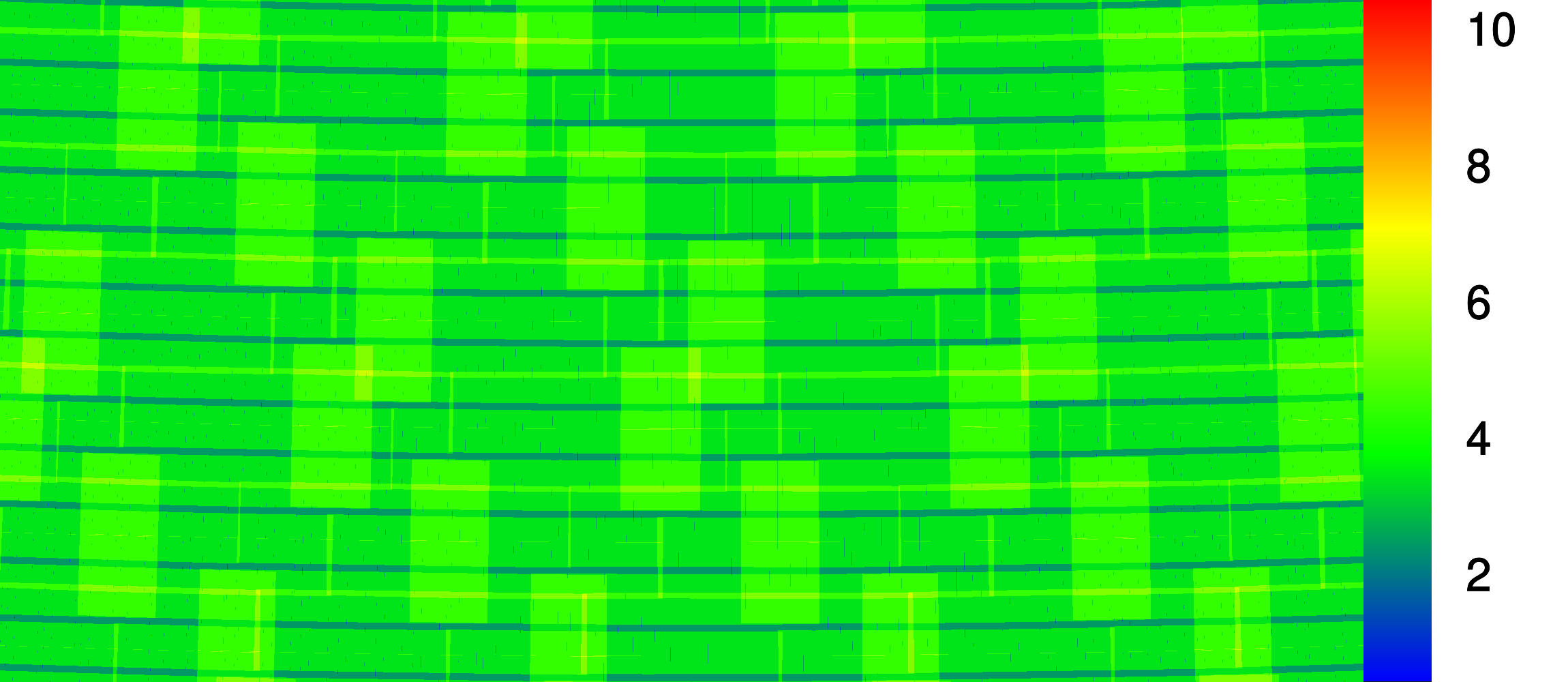}
  \end{center}
  \caption{Localized  spatial coverage in a $2^\circ\times 4^\circ$ area for the CFIS $u$-band (top panel) and CFIS $r$-band (bottom panel) data showing the observing pattern. The color scales are the same as in Figure~\ref{fig:coverage}.
  }
\label{fig:texture}
\end{figure}

Figure~\ref{fig:texture} shows the observing patterns for the CFIS $u$ and $r$ bands (top and bottom panels, respectively). At any given location, the goal is three exposures, which is obtained in practice for the majority of the sky. Due to the chip gaps in the MegaCam focal-plane array, there are small areas covered by only two exposures. Given the shape of the Megacam focal-plane, the areas covered by the “wing” chips overlap neighboring pointings and are consequently covered by more exposures.

\subsubsection{Data reduction} 
\label{sssec:CFIS.reduction}

CFIS data processing uses the MegaPipe pipeline \citep{2008PASP..120..212G}. The data reduction process starts with the raw images from CFHT. These are detrended with the Pitcairn software package, which performs much the same tasks as Elixir \citep{2004PASP..116..449M}. The overscan value is removed as usual and the bias, computed once per semester, is subtracted. The flat fields are generated once per dark run. For the $r$ band the flat fields are generated using the night-time science images themselves. Typically, there are several hundred $r$-band images taken in a run. The 100 longest exposures are combined using a median to produce a night sky flat. This approach was used in processing the data from the {\it Outer Solar System Origins Survey} \citep[OSSOS;][]{2018ApJS..236...18B} and was shown to produce slightly ($1-2$\%) deeper output images than using twilight flats. However, the $u$-band images do not contain enough sky photons to generate a satisfactory flat, and so twilight flats are used.

For the astrometric calibration, the first step is to run SExtractor (\citealt{bertin1996}) on each image. The parameters are set so as to extract only the most reliable objects ($5\sigma$ detections in at least five contiguous pixels). This catalog is cleaned of cosmic rays and extended objects using cuts on the half-light radius to leave only real objects with well-defined centers (stars and, to some degree, compact galaxies).

This observed catalog is matched to the astrometric reference catalog. The $(x,y)$ coordinates of the observed catalog are converted to (RA, Dec) using the initial world coordinate system (WCS) provided by CFHT. The match is done for each of the 40 MegaCam CCDs separately. The catalogs are shifted in RA and Dec with respect to one another until the best match between the two catalogs is found. If there is no good match for a particular CCD (for example when the initial WCS is erroneous), its WCS is replaced with a default WCS and the matching procedure is restarted. Typically 20 to 50 sources per CCD are found using this initial matching process. 

Higher-order astrometric terms are determined on the scale of the entire mosaic to allow the distortion of the full focal-plane to be measured. This distortion is well described by a polynomial with second- and fourth-order terms in radius measured from the center of the mosaic. While the distortion does not change drastically over time, it must be recomputed for every exposure. Measuring the distortion in this way means that only two parameters need to be determined (the coefficients of $r^2$ and $r^4$). Given that there are typically $20-50$ stars per chip, and 40 chips, this results in around 1000 data points. Less satisfactory is an analysis that is done chip-by-chip since a third-order solution requires 20 parameters per chip. For 40 chips, this is 800 parameters, and can lead to overfitting.

The astrometric reference catalog that we use is {\it Gaia} DR3 \citep{2023A&A...674A...1G}.  The positions of the reference sources are corrected for proper motion from the {\it Gaia} epoch (2016.0) to the epoch the image was taken. The images are astrometrically calibrated completely independently from each other. The astrometry is checked by examining the residuals with respect to {\it Gaia} and by cross-checking the stellar positions in overlapping images. Using this method, we find that the typical astrometric uncertainty is 20 mas.

The CFIS $r$-band photometry is calibrated using Pan-STARRS \citep{2020ApJS..251....6M} as a reference. The Pan-STARRS photometry is converted to the MegaCam system using the following transformation:

\begin{equation}
    r_{\rm Mega}-r_{\rm PS} = 0.003-0.050 x+0.0125 x^2-0.00699 x^3,
\end{equation}
where
\begin{equation}
   x=g_{\rm PS}-i_{\rm PS}~.
\end{equation}

\noindent This transformation is derived by computing synthetic photometry using the MegaCam and Pan-STARRS bandpasses multiplied by the Pickles \citep{1998PASP..110..863P} and CALSPEC spectra \citep{2020AJ....160...21B}.

Photometric calibration is done in two stages. The first stage deals with the variations of the zeropoint across the MegaPrime focal-plane. The illumination patterns shows slight variations from run to run and so are corrected on that timescale via the creation of a “super-flat”. For each image taken within a run, the mean zeropoint across the image is computed and the deviations relative to this mean are mapped as a function of CCD and position ($x, y$) within each CCD. The deviations are aggregated on a grid of $4 \times 9$ super-pixels and the median deviation is computed. Generally, the zeropoint offsets follow a consistent pattern from run to run, but there is some evolution over time. In a typical run, there will be several hundred images taken in each band, although sometimes a few runs have a smaller number of images, where there may be not enough images to produce an accurate mapping of the zeropoint variation. In these cases, the two neighboring runs (previous and following) are averaged to produce a map for the affected runs.

\begin{figure}
  \begin{center}
      \includegraphics[width=0.95\hsize]{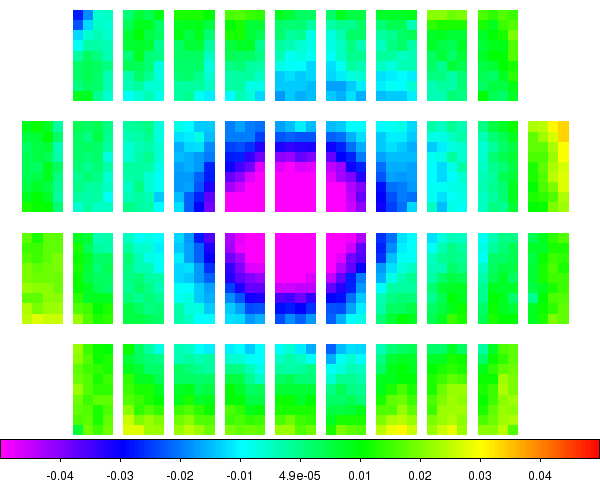}
  \end{center}
  \caption{Representative photometric super-flat for CFIS $r$-band data. The figure shows the variation of the zerpoint, measured in magnitudes, across the MegaCam focal plane.}
\label{fig:cfisSuperFlat}
\end{figure}

After the zero point variations have been mapped, they are removed by first expanding the $4 \times 9$ super-pixels to the original resolution of the CCDs, then converting the zeropoint difference (measured in magnitudes) to a flux ratio and multiplying each image by the result. The differential zeropoint is then remeasured on the corrected CFIS images and examined for large variations. Generally, this photometric super-flat correction is good to 1 mmag (0.1\%). There are a small number of exceptions on certain runs and certain CCDs that are on the order of 5 mmag in the $r$ band and 10 mmag in the $u$ band. A representative example of the super-flat correction is shown in Figure~\ref{fig:cfisSuperFlat}.

Once the differential photometric correction has been applied, an absolute zeropoint must be determined for the image as a whole. For the $r$ band, the zeropoint is determined by transforming the Pan-STARRS photometry into the MegaCam system, as before, to create a set of in-field standards. Typically, a few thousand stars are used to set the zeropoint. The $r$ band Pan-STARRS-to-MegaCam transformation is robust, given that the Pan-STARRS photometry is uniformly excellent, and Pan-STARRS covers the entire CFIS area. The $r$-band zeropoints are correspondingly accurate. The photometric residuals with respect to Pan-STARRS are typically 5\,mmag. Similarly, comparing the photometry of the indvidiual independently-calibrated images shows 5\,mmag residuals.

{\it Gaia} spectra \citep{2023A&A...674A...2D} form the basis of the $u$-band calibration.  The filter sensitivity function of the $u$ filter is well known, as discussed earlier in Section~\ref{sec:filters}. Multiplying the filter sensitivity function by the {\it Gaia} spectra results in synthetic $u$-band photometric standards over the entire sky, which we use to calibrate each MegaCam image. The major caveat on this method is that only relatively bright stars have {\it Gaia} spectra with reasonable signal-to-noise ratios. Consequently, it is important to reject potential calibration stars that are saturated in the MegaCam images to avoid erroneous solutions. The bright end saturation limit is around $u\sim14$ but varies by up to a magnitude with seeing (better seeing images saturate at fainter magnitudes. At the faint end, the {\it Gaia} spectra become unusable at $u\sim17$. A photometric comparison with the SDSS, and cross-comparison between individual images indicate a typical photometric precision of 12\,mmag.

The calibrated individual images are combined using SWarp \citep{2010ascl.soft10068B} and  resampled on to a set of tiles for both $u$ and $r$ band covering the sky. The tiles are $10000 \times 10000$ pixels with a pixel size of 0.1857$^{\prime\prime}$, approximately matching the native resolution of MegaCam. The tiles are spaced exactly 0.5$^\circ$ apart in declination and $0.5/\cos(\delta)$ apart in right ascension. Adjacent tiles overlap each other by approximately 3\%. SWarp resamples the input images based on their astrometric calibration and scales the images according to the photometric calibration. The resampled images are combined with a weighted mean. Weight maps are generated with WeightWatcher \citep{2008ASPC..394..619M}, which masks cosmic rays, bad columns, and other CCD defects. 

Two sets of stacks are produced for the $r$-band images. The first uses SWarp's background estimator as described above, which is reliable, but removes any extended emission beyond the scale of the background mesh (128 pixels). Therefore, a second set of low surface brightness stacks are also created, using the Elixir-LSB methodology described in \cite{2012ApJS..200....4F} and mentioned in the previous subsection. In brief, the sky background for an image is measured from other images taken adjacent in time in the observing sequence. The images are median-filtered to remove small bright sources, scaled to a common average background level, and then combined using an aggressive sigma-clipping scheme. 

\subsection{Pan-STARRS} 

The Panoramic Survey Telescope and Rapid Response System \citep[Pan-STARRS;][]{2002SPIE.4836..154K} currently consists of two telescopes, PS1 and PS2, making up the Pan-STARRS Observatories. PS1 has been in operation since the later part of 2009, and PS2 began observing in 2015, with full normal operations in 2017. Each telescope has a 1.8m primary mirror, a 0.9m secondary mirror,
and a large camera mounted at the Cassegrain focus behind a corrector lens system, pneumatic filter changer, and shutter. 
The optical designs of both PS1 and PS2 are identical; however, the telescopes and domes were built by different manufacturers and are quite different. 
The Gigapixel Camera 1 (GPC1) and Gigapixel Camera 2 (GPC2) both have orthogonal transfer arrays, but are operated as normal CCDs. 
GPC2 has less correlated read noise and better charge-transfer properties, but more problematic devices that create gaps in the focal-plane. The filter coating specifications for both cameras are identical. 
The Pan-STARRS1 magnitude system is very well defined, including both the instrumental sensitivity and atmospheric transmission functions \citep{2012ApJ...750...99T}

In addition to the observatories, Pan-STARRS has substantial computing capability and highly developed software systems, including the Image Processing Pipeline
\citep[IPP,][]{2020ApJS..251....3M}
and the Moving Object Processing System
\citep[MOPS,][]{2013PASP..125..357D},
necessary to reduce and analyze the nightly flow of the time-sensitive data, as well as periodic re-processing, that produces stacked images and other reduced data products.

\subsubsection{Survey strategy} 

The Pan-STARRS contribution to the UNIONS data set includes the combined total historical $i$-band data taken with both telescopes in the UNIONS footprint and the $z$-band data in the range $30 \le \delta \le 42^\circ$ within the UNIONS footprint.
From September 2009 to April 2014, PS1 data were taken under the auspices of the Pan-STARRS1 Science Consortium, resulting in the Pan-STARRS1 Surveys \citep{2016arXiv161205560C}.
The strategy for the $3\pi$ Survey (all sky north of $\delta = -30^\circ$, or 3$\pi$ steradians) was focused on obtaining uniform coverage over the $3\pi$ area in five bands while meeting multiple science goals. The result was a complicated algorithm designed to; (i) make best use of twilight; (ii) measure parallaxes for as many stars as possible (in the $y$ and $z$ bands); (iii) make multiple exposures separated by $\sim 20$ minutes to discover moving objects at opposition; (iv) return as soon as possible to recover transient objects; and (v) cover the sky in every band as nearly uniformly as possible given the weather. We refer readers to \cite{2016arXiv161205560C} for more details. 

Since 2015, Pan-STARRS operations have been funded at the 90\% level by the NASA Near Earth Object (NEO) program, with the primary goal of discovering NEOs greater than 140m in diameter. As such, the survey is optimized for this goal, not for the purpose of obtaining uniform sky coverage. However, searching for NEOs in the Galactic plane with our technique of difference imaging has proven to be less than satisfactory, so the NEO survey has concentrated on the extragalactic sky. The data from Pan-STARRS that are included in the UNIONS survey are restricted to the data taken that lie within the main UNIONS footprint, although we intend to release all the data eventually. The NEO survey is carried out in 
$w$ band (a wide filter that is equivalent to $g+r+i$) during dark time, and $i$ band in bright time with four nights centered around full moon in $z$-band. 
For more details on the NEO survey see 
\citet{2016IAUS..318..293W}
and
\citet{2022AERO53065.9843625}.
Since the successful addition of the WISHES program into UNIONS, the $z$-band coverage has largely stopped, with the full moon time reverting to $i$ band. 

Both survey strategies use approximately 45\,s exposures (it has varied slightly over time). A consequence of this is the very large number of exposures necessary to reach the required total exposure time of 20,000\,s per pixel, or about 450 overlapping raw images.  After warping and registration, the number of reduced images that can contribute to a given stack can be much higher, since many will only be partially populated. The large amount of masking due to the cosmetics and charge transfer issues in the Pan-STARRS cameras, together with the very large number of images that go into a stack, means that the number of valid pixels in a given stack varies greatly from pixel to pixel. This also means that the data volume from the Pan-STARRS $i$-band and $z$-band contributions is two orders of magnitude greater than in the other bands from other surveys. 

\begin{figure}
  \begin{center}
      \includegraphics[width=0.95\hsize]{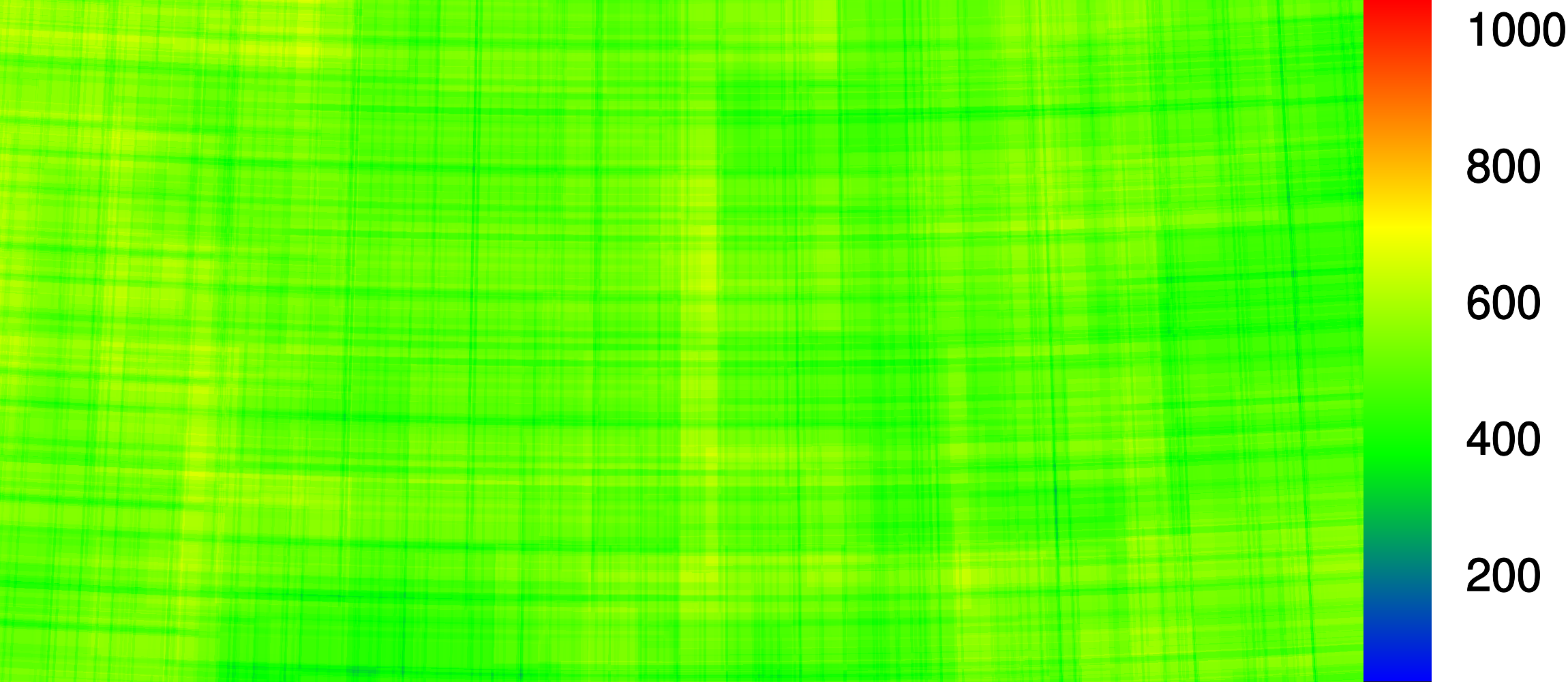}
      \includegraphics[width=0.95\hsize]{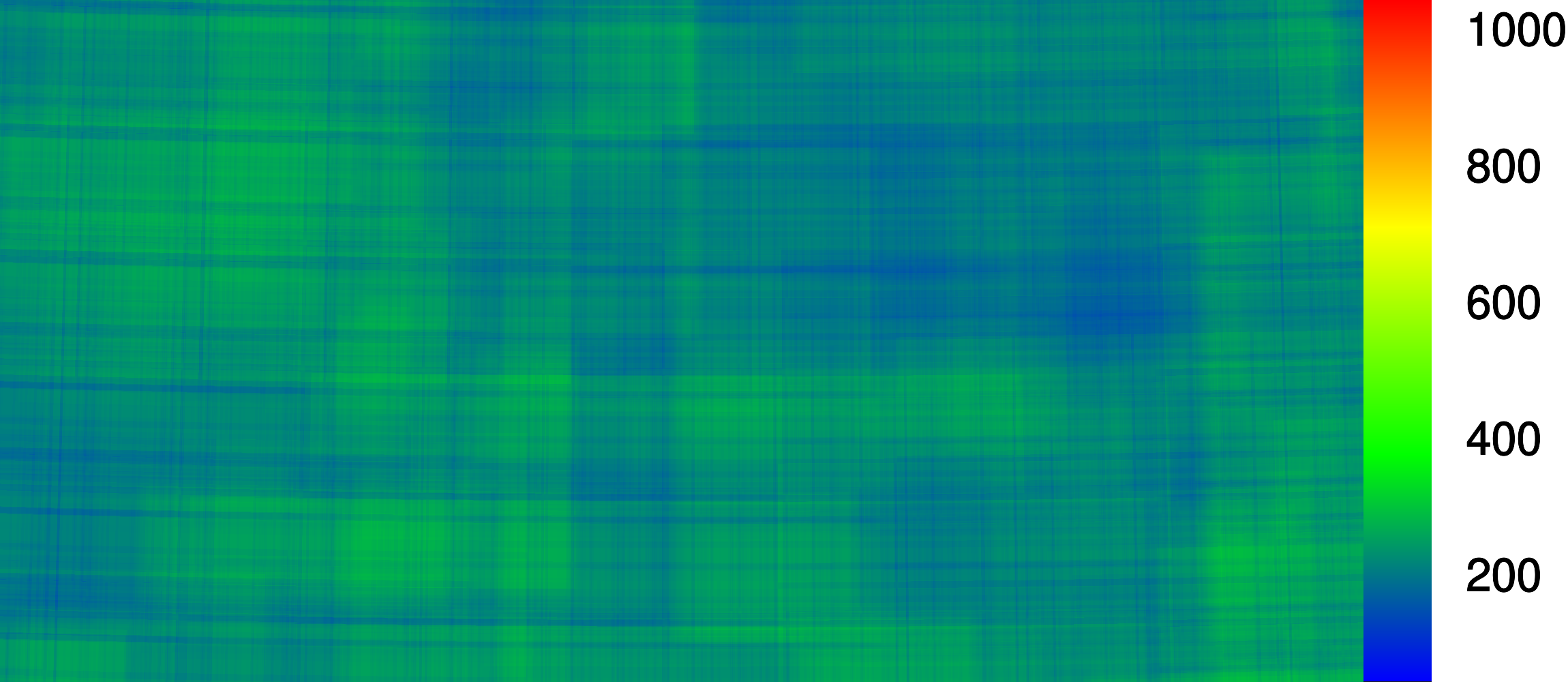}
  \end{center}
  \caption{Localized spatial coverage in a $2^\circ\times 4^\circ$ area for the Pan-STARRS $i$-band (top panel) and $z$-band data (bottom panel) showing the observing pattern. The color scales are the same as in Figure~\ref{fig:coverage}.
  }
\label{fig:pstexture}
\end{figure}

The top and bottom panels of Figure \ref{fig:pstexture} show the observing pattern for the UNIONS Pan-STARRS data in $i$ and $z$, respectively. The significantly larger number of exposures acquired by Pan-STARRS means that the coverage is correspondingly more uniform, relative to the CFHT and Subaru components of UNIONS.

\subsubsection{Data reduction} 

Pan-STARRS images are processed using the Pan-STARRS IPP and combined into deep stacks using a modification of the stacking analysis described by \cite{2020ApJS..251....4W}, as discussed below.  
Since the Pan-STARRS observations for any portion of the UNIONS footprint accumulate over many years, we have had the opportunity to produce stacks at multiple dates with the data available up to that point.  As of January 2025, we have generated four sets of deep stack images, identified in what follows as the UNIONS Pan-STARRS (Internal) Data Releases 1-4.  Since these have been used over the years by members of the UNIONS consortium for science results and are available as part of the publicly released data, we include a discussion of these stacks and their differences in the following.


The UNIONS Pan-STARRS DR1 used observations from 5 of the 10 Pan-STARRS\,1 Science Consortium (PS1SC) Medium Deep (MD) Survey fields to generate stacks with the expected depth from the full UNIONS survey in each of the $g,r,i$ and $z$ bands.  These observations were taken during the PS1SC mission between January 2010 and April 2014.  The PS1SC MD fields were observed many times over the roughly 4 years of the PS1SC mission, with typical exposures times of [$g,r,i,z$] = [113, 113, 240, 240] seconds.  Within a given night, one or two filters were used for observations with a rotational dither sequence.  A series of eight exposures was obtained for each sequence with the telescope boresite held fixed on the sky and the position angle of the camera rotated through 120\degrees\ in steps of 15\degrees.  The resulting circular 7\,deg$^2$ field of view was thus well covered, with detector gaps largely filled in for each sequence.  Each field has roughly 800 such exposures per filter, which amounts to far more exposure time than needed to reach the UNIONS depth goals.  To approximate the expected depth, we selected a randomly-drawn subset of exposures, with total exposure times of [$g,r,i,z$] $=$ [30,000, 15,000, 12,500, 12,500] seconds, i.e., roughly [270, 130, 50, 50] exposures.  Exposures were randomly drawn from those available in order to give a somewhat realistic distribution of seeing, sky background, and transparency.  

These exposures were reprocessed from the raw observations following the process described by \cite{2020ApJS..251....3M}, with the astrometric and photometric calibration tied to the Pan-STARRS $3\pi$ DR2 analysis \citep{2020ApJS..251....6M}.  Astrometry is thus tied to the {\it Gaia} DR1 system through this calibration, without correction for proper motions.  Individual exposures were warped to a common pixel grid defined to mimic the CFIS tiling of the sky: projections 0.5 degree on a side with $10 000 \times 10 000$ pixels (as described in ection~\ref{sssec:CFIS.reduction}).  
The warps were combined to generate stacks using the stacking procedure described by \cite{2020ApJS..251....4W}. 



The UNIONS Pan-STARRS DR2 used observations in $i$ and $z$ band covering two regions totaling approximately 1000~deg$^2$.  The release was generated in Summer 2019 using data obtained through April 2019.  The North Galactic Cap portion covers $8^{\rm{h}} \le {\rm RA} \le 12^{\rm{h}}$, $30^\circ \le \delta \le 40^\circ$. The South Galactic Cap portion covers $23^{\rm{h}} \le {\rm RA} \le 3^{\rm{h}}$, $30^\circ \le \delta \le 40^\circ$. Total exposure times in the regions range from 3000 to  10\,000 seconds in $i$ band, and roughly 10\,000 seconds for most of the area in $z$ band.  These exposures were again processed from the raw data to the full stacks, as was done for the UNIONS Pan-STARRS DR1 data, using the same CFIS tessellation discussed above.

The UNIONS Pan-STARRS DR3 used observations in only $i$ band covering the full UNIONS North Galactic Cap region, roughly 5000~deg$^2$.   This analysis was performed in late 2021 and early 2022 using observations obtained through November 2021.  For this analysis, we bypassed the large amount of computation time needed to reprocess the observations from raw images by starting with data already processed to the warp stage.  The Pan-STARRS team maintains an archive of all observations obtained to date, processed to the warp stage for rapid asteroid recovery.  A particular exposure may have been processed through the IPP multiple times over the years, as the analysis system has been improved, but we do not have sufficient storage to maintain multiple copies of the processed warp images.  Thus, only a single version is available on the IPP archive, with some heterogeneity in the software versions used.  For the DR3 stacks, we used the warps generated for the Pan-STARRS $3\pi$ Survey DR2 release for exposures obtained before early 2015.  These early exposures are thus tied to the The Two Micron All Sky Survey \citep[2MASS;][]{2006AJ....131.1163S} astrometric reference frame used for the pixel-level analysis of the $3\pi$ survey data; the catalog released by the Mikulski Archive for Space Telescopes (MAST) in 2019 was tied to {\it Gaia} DR1 in the database after the pixels had been processed.  For subsequent exposures, we used the warps generated by the nightly processing system.  Until 19 September 2017, these exposures used the same reference catalog used for the $3\pi$ survey, astrometrically tied to 2MASS.  After that date, the nightly processing system used the internal reference catalog based on the $3\pi$ release data, with the astrometry tied to {\it Gaia} DR1, until 27 April 2023.  Starting on that date the astrometry was tied to {\it Gaia} EDR3, including compensation for proper motion to the epoch of the observations.  Note that the stellar images in the stacks generated over this large time baseline will be affected by the proper motion of the stars.  Typical stellar proper motions ($\sim 5$ milliarcseconds per year) are equivalent to roughly a quarter of a pixel, while moderately high proper motion stars may be smeared by 2 or 3 pixels.

The warps use the standard Pan-STARRS tiling scheme described by \cite{2020ApJS..251....3M} called ``RINGS.V3''.  In this scheme, warps are roughly 6600$\times$6600 pixels, with 0.25$^\prime\prime$ pixels yielding tiles with 27.5 arcmin on a side and with an approximately 1 arcminute overlap on the edges.  The stacking process used for this analysis differs somewhat from that described by \cite{2020ApJS..251....4W}.  That analysis generated a stack by convolving the input warps with a smoothing kernel chosen to result in the same target PSF for the entire input set.  The routine also generates an unconvolved version of the stack.  The goal of this PSF-matching process was twofold.  First, it was meant to enable photometry of the stack with a well-defined PSF, but the results of this technique have been unsatisfactory  \citep[see discussion in][]{2020ApJS..251....5M}.  The other purpose of the convolution was to allow the rejection of outlier pixels with a consistent PSF to avoid inconsistent rejection of pixels in the cores of stellar images, again to avoid disturbing the PSF shapes.  The concerns about inconsistent pixel rejections have proven unimportant for stacks using hundreds of input images.  Given the high computational cost of the convolutions used previously, we have modified our stacking process to eliminate that step, and only generate an unconvolved stack image in our analysis.  With the large number of input exposures, we are able to more robustly reject outlier pixels by sorting the input values (for a given pixel) and calculating the average after rejecting a fixed (symmetric) fraction of these values.  Note that this calculation is similar to the calculation of the median, but results in lower variance than the median. 


The UNIONS Pan-STARRS DR4 data set used observations from both Pan-STARRS\,1 and Pan-STARRS\,2 telescopes to generate deep stacks in $i$ band and $z$ band.  The $i$-band stacks cover the UNIONS North Galactic Cap region and the $z$-band stacks cover the region $8^{\rm{h}} \le {\rm RA} \le 18^{\rm{h}}$, $30 \le \delta \le 42^\circ$.  The $i$-band stacks were generated in Summer 2023 using observations obtained through June 2023.  The $z$-band stacks were generated in February 2024 using observations available through that month.  The PS1 and PS2 stacks were generated independently, then combined together using a simple average, weighted by per-pixel inverse variance. 

The input warp images are photometrically calibrated by the nightly science analysis system and tied to the Pan-STARRS $3\pi$ Survey DR2 release.  The Pan-STARRS stacks have highly textured point spread functions from the many input images, making accurate PSF photometry measurements directly from the stack pixels challenging.  We have performed demonstration photometric analysis tests using the stack positions to force the PSF photometry on the input warp images.  These show that deep photometry may be measured with a bright-end RMS scatter of approximately 5 to 8\,mmag in comparison with the Pan-STARRS $3\pi$ Survey photometry. 


\subsection{WISHES}

WISHES is an open-use program of the Subaru Telescope approved in June 2020 as an Intensive Program using the HSC instrument, a wide-field imager mounted on the prime focus of the Subaru 8.2\,m telescope \citep{2018PASJ...70S...1M}. HSC uses 104 science CCDs to cover a $1.5^\circ$ diameter field of view with a pixel size of $0.168^{\prime\prime}$ (\citealt{2018PASJ...70S...2K}; see also \citealt{2018PASJ...70S...1M} and \citealt{2018PASJ...70S...4A}). The properties of the broad-band $grizy$ filters are presented in \citet{2018PASJ...70...66K} and summarized in Section~\ref{sec:filters}. Thanks to carefully designed optics, HSC achieves seeing-limited imaging on Maunakea during most of its observations \citep{2018PASJ...70S...1M,2018PASJ...70S...3F}. A second Subaru Intensive Program, WISHES+, was accepted for observing between S24B and S27A, to survey the NGC region below $\delta = 30^\circ$ down to $\delta = 15^\circ$ as outlined in Figure~1.

\subsubsection{Survey strategy} 

WISHES obtains the majority of the $z$-band imaging for UNIONS, and complements the $z$-band observations previously obtained by Pan-STARRS. The exposure time of each HSC pointing is set to 240\,s, which is split into three dithered sub-exposures. The target limiting magnitude is $23.4$\,mag ($10\sigma$ within a $2^{\prime\prime}$ aperture). This is deeper than the other large $z$-band surveys in this region of sky, the Mayall $z$-band Legacy Survey \citep[MzLS;][]{2018PASP..130h5001Z,2019AJ....157..168D} and Pan-STARRS 3$\pi$, by approximately 1 and 2~mag, respectively. The total on-source observing time (excluding readout and other overhead time) of WISHES for the original area around the NGC is 200.9 hours, which roughly corresponds to 40 nights, including overheads and typical weather. WISHES+ corresponds to an additional 106 hours (15.1 nights).

We searched for the archival HSC data \citep{2002ASPC..281..298B} to avoid duplications of HSC $z$-band imaging in the WISHES footprint. We find that the Hawai'i eROSITA Ecliptic Pole Survey \citep[HEROES;][]{2018ApJ...859...91S,2020ApJ...895..132T,2021ApJ...914...79T} obtained deep $z$-band imaging data over approximately 45\,deg$^2$ of the sky centered at the northern ecliptic pole. These images typically have a seeing of $0.6$\,arcsec and achieve a depth of 24.2~mag ($10\sigma$ within a $2^{\prime\prime}$ diameter aperture), which is of sufficient quality to be used for WISHES. In addition, the HSC Subaru Strategic Program \citep[HSC-SSP;][]{2018PASJ...70S...4A} Wide layer targets an approximately 50\,deg$^2$ region at $\delta \simeq 43^\circ$ and $200 \lesssim {\rm RA} \lesssim 250^\circ$. Given the total exposure time of 1200s for each HSC pointing, these images are much deeper than WISHES. We therefore removed from WISHES any fields that overlap significantly with HEROES and the HSC-SSP Wide layer. 

Given the challenges of tesselating the HSC footprint on the celestial sphere, we chose to follow the HSC-SSP \citep{2018PASJ...70S...4A} strategy, and allow for overlaps of the HSC pointings at the edges to ensure an efficient tiling pattern. This results in an effective area of each HSC pointing of 1.46\,deg$^2$ (the area of an inscribed regular hexagon in the HSC field of view). We used the method described in \citet{2006QJRMS.132.1769S} to calculate the locations of all the fields necessary to cover the WISHES footprint, resulting in a total of 3014 HSC pointings. For WISHES+, an additional 928 pointings are required. 

\begin{figure}
  \begin{center}
      \includegraphics[width=0.95\hsize]{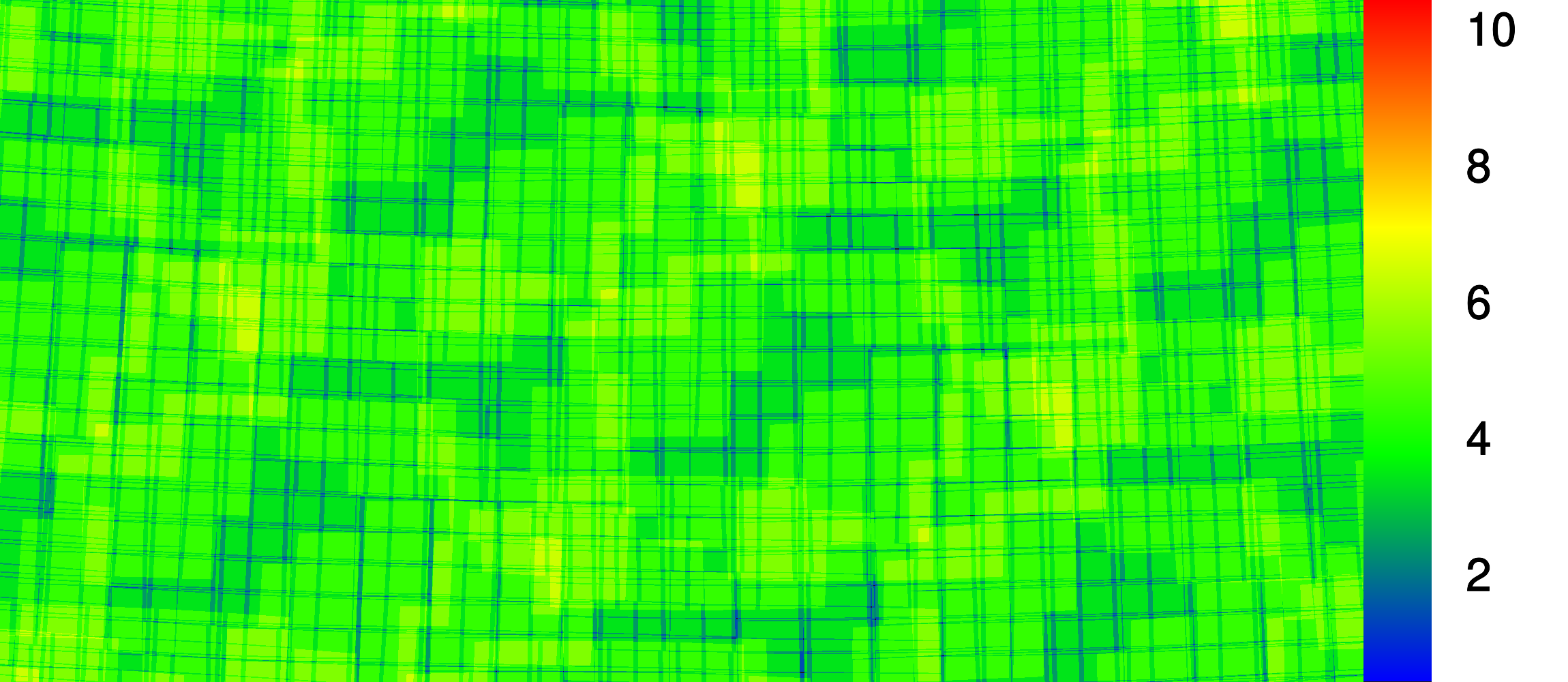}
  \end{center}
  \caption{Localized spatial coverage in a $2^\circ\times 4^\circ$ area for the WISHES $z$-band data showing the observing pattern. The color scales are the same as in Figure~\ref{fig:coverage}.
  }
\label{fig:hscztexture}
\end{figure}

Following the dithering strategy in the HSC-SSP Wide layer \citep{2018PASJ...70S...4A}, we adopt a dithering pattern with large offsets to mitigate an inhomogeneous depth due to vignetting. Specifically, for each pointing center, we obtain three dithered exposures that are located on a circle of radius 0.3$^\circ$ centered at that position, such that the three positions are equally spaced on the circle and the first position is rotated clockwise by 21$^\circ$ from the North-South direction (i.e., {\tt NDITH} =  3, {\tt RDITH} = 0.3$^\circ$, and {\tt TDITH} = 21$^\circ$).  Figure \ref{fig:hscztexture} illustrates the WISHES observing pattern.

All the WISHES observations are conducted in queue mode. The requested observing conditions are that the seeing is better than 1$^{\prime\prime}$, the transparency is better than $0.7$, and the airmass is less than $2.0$. Observations are conducted on dark and gray nights with a requirement that the target field should be at least 30$^\circ$ away from the Moon. There is no requirement on the cadence of our observations.

\subsubsection{Data reduction} 
The data processing for WISHES is performed with a variant of the LSST Science Pipeline called hscPipe version 8.5.3. This is exactly the same version of the software that was used in the HSC Subaru Strategic Program (SSP) survey \citep{2022PASJ...74..247A} and we follow the same process. Briefly, all CCDs are first detrended and calibrated for astrometry and photometry using Pan-STARRS DR1 \citep{2020ApJS..251....6M} as the reference catalog. No fringe correction is applied. Second, joint-exposure calibrations are performed to refine the calibrations on individual CCDs for both astrometry and photometry. Then, calibrated CCDs are assembled and combined into coadded images on pre-defined tiles. In the coaddition process, only the CCD images that satisfy a condition of seeing $<1.3^{\prime\prime}$, transparency $>0.3$, and sky background count $<45000$ DN are included. The global sky background pattern is subtracted so that the resultant images preserve light profiles of extended or bright objects. 
Finally, sources are detected and measured with various algorithms on the coadded images. In this catalog generation step, object peaks are first detected based on the S/N ratio of the objects, and fluxes belonging to individual peaks are deblended into separate footprints. Then, photometric parameters are measured for each object. The flux measurement algorithms include fixed-aperture, Kron, PSF-fit, and CModel (a galaxy model fit combining an exponential disk and a de~Vaucouleurs profile) photometry.

A major difference of the WISHES processing from that adopted in the SSP is the configuration of the sky tiles across the survey footprint. In the SSP, they are based on the ring tiles with approximately 1.7$\circ$ widths, and each tile (or ``tract'', in the LSST terminology) was further divided into 81 sub-areas (called patches) for processing. However, we adopt the CFIS tiling strategy to make the co-added images have the same geometry as the other survey components. In addition, because we process only $z$-band images, we do not apply the {\it FGCM} \citep{2018AJ....155...41B} calibration, which is designed to deliver better calibration when applied on multiple bands. We instead use {\it jointcal} \citep{jointcal}  for both astrometry and photometry. The resulting photometric calibration is comparable to the that of the Hyper Suprime-Cam Subaru Strategic Program \citep{2022PASJ...74..247A} $i$-band data, with a typical zeropoint uncertainty of 5\,mmag.

\subsection{WHIGS}

WHIGS, like WISHES, uses the HSC on Subaru. However, observing time is obtained through a combination of a Canadian-led Subaru Intensive Program via Gemini-Subaru exchange time (led by MH) and through PI programs led from the University of Hawai'i (led by KC).

\subsubsection{Survey strategy} 

\begin{figure}
  \begin{center}
      \includegraphics[width=0.95\hsize]{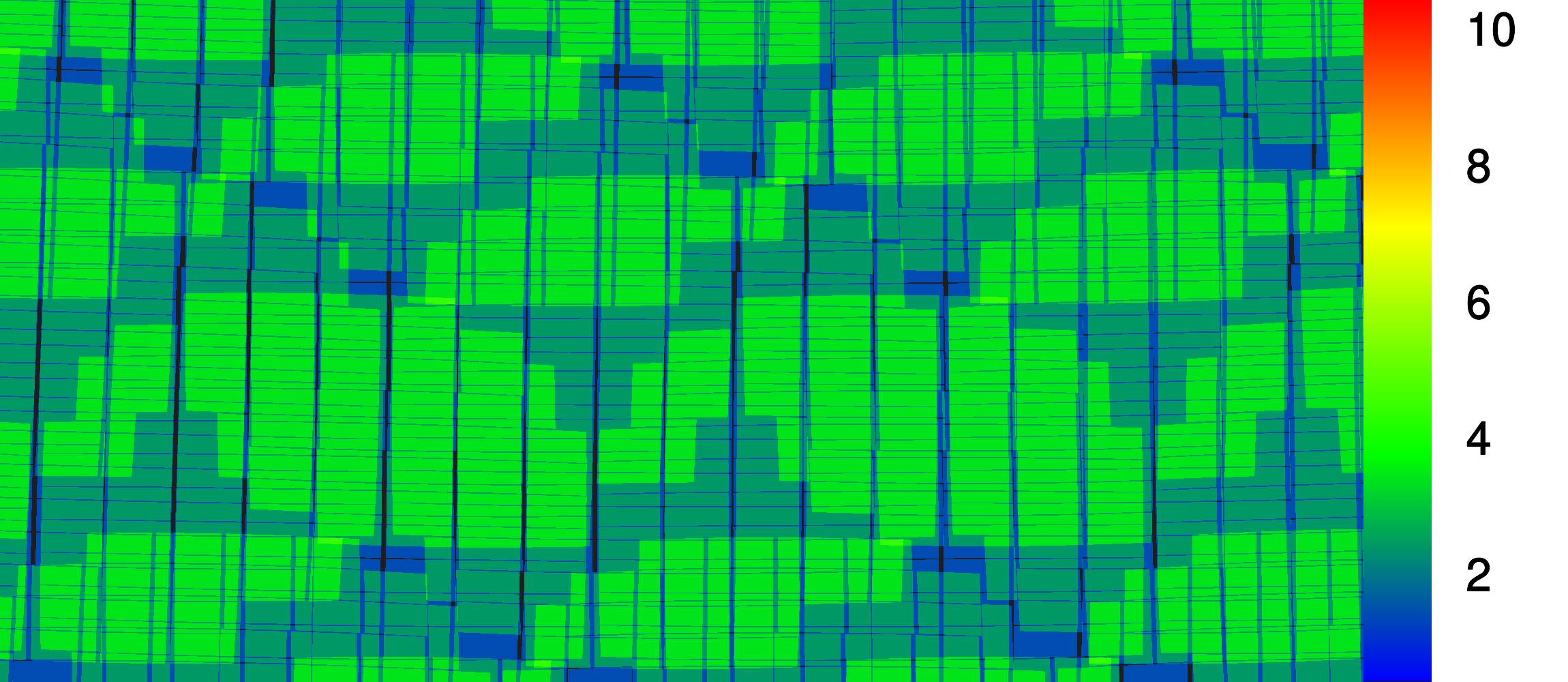}
  \end{center}
  \caption{The localized spatial coverage in a $2^\circ\times 4^\circ$ area for the WHIGS $g$-band showing the observing pattern. The color scales are the same as Figure~\ref{fig:coverage}
  }
\label{fig:hscgtexture}
\end{figure}

The observing strategy of WHIGS is designed to work around existing archival $g$-band data, resulting in a heterogeneous tiling scheme that optimizes the area covered to an adequate depth, in preference to ensuring that small gaps are covered. We use three different sky tessellations that, like WISHES, treat the field of view of HSC as hexagonal. These different tessellations serve as dithers that fill the gaps between chips and enable bad-pixel rejection, and result in an effective field of view for tiling of 1.46~deg$^2$. Figure \ref{fig:hscgtexture} illustrates the WHIGS observing pattern.

Our nominal integration time is 150s, split as three subexposures of 50\,s each. Including overheads of 40\,s per subexposure results in a total time of 270\,s per HSC field of view.  Observing was conducted primarily in classical mode, and so the nominal exposure time of 50\,s is adjusted in real-time to account for changes in weather conditions and observing transparency. Our baseline conditions assume a transparency of 0.8, at least 4 days from full Moon with a 65$^\circ$ Moon distance. As described in Section~\ref{sec:filters}, this allows us to reach our target limiting magnitude of $g = 24.5~(10\,\sigma, 2^{\prime\prime}$ diameter aperture).

\subsubsection{Data reduction} 

Processing of the WHIGS data, like WISHES, is done using the LSST Science Pipeline \citep{2018PASJ...70S...5B, 2019ASPC..523..521B}. Images are detrended to correct for bias, dark current, and flat-field effects. Given that WHIGS is limited to the $g$-band filter, there is no need to perform fringe corrections. Local sky backgrounds are determined and subtracted internally using a grid of 128 pixel bins across the focal-plane. Subsequently, astrometric calibration of individual images is performed against the {\it Gaia} DR2 catalog \citep{2016A&A...595A...1G, 2018A&A...616A...1G} and photometric calibration against the Pan-STARRS DR1 catalog \citep{2020ApJS..251....6M}. The calibrated final photometry is in the HSC filter system and in AB magnitudes.

Following calibration of the individual images, co-adds are made on the same tile grid as the CFIS survey. For each tile, only individual images with a S/N $> 5$ at our reference magnitude of $g = 24.5$ are included in the stacks. This results in a total of 20644 tiles with $g$-band coverage as of January 2025. Object detection on the co-added images is performed using the default setting of the LSST Science Pipelines peak-finding algorithm, which discards sources with S/N $\le 5$. 

Deblending of sources is performed to remove the light of neighboring, blended objects when finding and measuring sources. We make use of the Scarlet deblender \citep{2018A&C....24..129M}, which is available to use within the LSST Pipelines framework, and produces better results than the default option at the same computational resource cost. Following this, PSF-fitting photometry is performed on the object list to produce the final deep co-added catalog, producing positions, magnitudes, and star-galaxy separation metrics. Again, the resulting photometric calibration is comparable to the that of the Hyper Suprime-Cam Subaru Strategic Program $g$-band data with a typical zeropoint uncertainty of 6\,mmag.



\section{Unified multiband catalogs} 
\label{sec:guc}

Given the nature of UNIONS as a multi-telescope survey, the survey components have proceeded at different rates. Because the unique constraints at each telescope specific to the time allocations, different regions of the main survey area were observed at different times. As such, the overlap between the different bands was initially quite limited and it made sense for the photometric catalogs to be created independently. Now, with full five-band coverage of the majority of the main survey area, homogeneous photometric catalogs are being created in which all five bands are processed simultaneously. Here, we describe the first of these catalogs, which uses the GAaP \citep[Gaussian Aperture and PSF;][]{kuijken2008,kuijken2015,kuijken2019} method, which will be made available to the community upon release of the UNIONS data set.

The GAaP method provides accurate high S/N multi-band photometry of the typical small galaxies (and stars) in high Galactic latitude wide-field imaging surveys. The processing involves a shapelet-based PSF model that is being used to define a spatially varying kernel, which convolves an image to a common (i.e., spatially constant), isotropic Gaussian PSF. This is done for each band individually, keeping each PSF as compact as possible and so avoiding any loss of information due to blurring. 

The resulting set of so-called "Gaussianized" images is then ideally suited for forced photometry with Gaussian-weighted, elliptical apertures. The Gaussian weight function is chosen individually for each band to account for the PSF size differences between bands, yielding accurate colors of the same physical parts of objects across wavelengths. Flux errors are estimated by taking into account the correlated noise introduced by the convolution and any other aspect of data reduction via a direct measurement of the pixel-to-pixel auto-correlation.

As an aperture-based forced photometry method, GAaP fluxes are generally not total fluxes, with light leaking out of the aperture for extended objects. For stars, however, GAaP yields accurate total fluxes across bands. The Gaussian weighting means that the central parts of galaxies are up-weighted in comparison to their outskirts. In the absence of color gradients within an object, this still yields accurate global colors, whereas otherwise, colors are biased towards the center. For a typical spiral galaxy, with a redder core and a bluer extended disk, this has the added benefit of concentrating on those central parts of the galaxy with a pronounced 4000\AA\ break making it easier to estimate a photometric redshift. However, the radial weighting might bias the global colour of an extended galaxy in presence of such colour gradients, which needs to be taken into account for some other applications.

Practically, GAaP is implemented such that it can easily work with images from different cameras, different footprints, and different pixel grids. The method can be applied to stacks as well as individual exposures, both of which modes have been used extensively in the Kilo-Degree Survey \citep[KiDS;][]{dejong2013}, especially in its combination with the VIKING (VISTA Kilo-degree INfrared Galaxy survey) infrared data \citep{wright2019,wright2024,kuijken2019}. 

For UNIONS, the GAaP method is applied to stacks from all available bands in $30'\times30'$ tiles using the versatile \textsc{PhotoPipe}\footnote{\url{https://github.com/hendrik1008/PhotoPipeUNIONS}} package \citep{wright2024}. These UNIONS tiles are arbitrarily defined and do not correspond to a particular pattern of observation. As such, exposures with quite different PSF properties can enter a stack. Careful selection of stars is performed to capture those PSF variations in the model, but it is clear that this aspect of the implementation ultimately limits the accuracy of the photometry.

A SExtractor \citep{bertin1996} catalog based on the CFIS $r$-band tiles is used to define the elliptical apertures for the forced photometry step on all five UNIONS bands. Subsequently, the stellar multi-band photometry is compared to an external reference catalog, such as SDSS, as a means of tile-wise quality control. 

The algorithms are optimized such that a five-band survey with the area of UNIONS can be run in approximately days on 100 modern CPU cores (depending somewhat on I/O limits and overheads). In the future, it is planned to apply GAaP to individual UNIONS exposures instead of stacks (similar to what has been done by the KiDS team with the VIKING infrared data), which should improve the PSF modeling further and enhance the robustness and S/N of the GAaP multi-band photometry at the expense of significantly higher computational demands.

\begin{figure}
  \begin{center}
      \includegraphics[clip=true, trim=0.5cm 0cm 1.5cm 0.5cm, width=0.95\hsize]{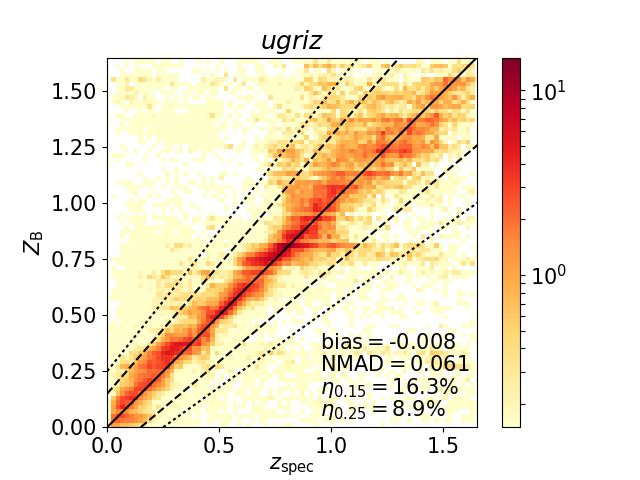}
  \end{center}
  \caption{Photometric redshifts estimated with BPZ from the GAaP catalog versus spectroscopic redshifts. This logarithmically-scaled density plot (arbitrarily normalised) shows the quality of $ugriz$ photo-$z$s obtained from UNIONS multi-band photometry. The dashed and dotted lines indicate the limits for outliers, as defined in Section \ref{sec:guc}.
  }
\label{fig:photoz}
\end{figure}

To demonstrate the utility of this catalog, Figure~\ref{fig:photoz} presents a comparison between photometric redshifts, calculated using the Bayesian Photometric Redshift code (BPZ; Benítez 2000), and a high-quality spectroscopic redshift compilation. This compilation was created by retrieving, cross-matching and merging major spectroscopic surveys including SDSS, DESI, HectoMAP \citep{2023ApJ...945...94S} within the UNIONS footprint, using the median redshift for objects with multiple measurements and retaining only redshifts with high quality flags. A forthcoming publication will provide a complete description of the spectroscopic data compilation. The spectroscopic sample has been weighted such that its $r$-band number-counts correspond to the UNIONS $r$-band number-counts in the GAaP catalog. Hence, the performance statistics quoted in Figure~\ref{fig:photoz} give a fair representation of what can be expected from the full UNIONS catalog. We compute the median and normalized median absolute deviation (NMAD) of the quantity 
\begin{equation}
\frac{\Delta z}{1+z}\equiv\frac{z_{\rm spec}-z_{\rm phot}}{1+z_{\rm spec}}
\end{equation} 
representing the photo-$z$ bias and scatter, respectively. We also quote the rate of outliers,  $\eta$, which we define as objects with $\Delta z/(1+z)>0.15$ (reported as $\eta_{0.15}$) and $\Delta z/(1+z)>0.25$ (reported as $\eta_{0.25}$). The photo-$z$ quality shown here compares favourably to previous stage-III imaging surveys \citep[e.g. compare to Fig.~7 of][]{wright2019} and already comes close to the Euclid primary science requirements for photo-$z$ ($\sim5\%$ scatter and $\sim10\%$ outliers) even before adding the space-based Euclid photometry to UNIONS, especially the three high-SNR NISP-P infrared bands.

\section{UNIONS and {\it Euclid}} 

While UNIONS is a scientific survey in its own right, it is also a major contributor to the ground-based component of the {\it Euclid} space mission, as discussed in Section \ref{sssec:unionseuclid}. The {\it Euclid} Consortium (EC) is producing its own advanced data products from UNIONS data, and the pipeline that is being used is different from the pipelines described in the previous sections. {\it Euclid} data processing is outlined in Sections 7.3 and 7.4 of \citet{euclid2024}, and as such {\it Euclid} data products based on UNIONS data should not be confused with the UNIONS data products -- including the processed images, stacks and unified photometric catalogs -- described in the previous sections of this paper.

Within the EC, OU-EXT (Organizational Unit: External) is responsible for the ``Euclidization'' of external surveys. OU-EXT takes as input the UNIONS Single Epoch Frames (SEFs). UNIONS delivers the SEFs fully detrended, with a full astrometric and phototometric solution. OU-EXT does not repeat the detrending, nor does it recompute the astrometric solution. However, it does redo the photometric calibration using {\it Gaia}. In Phase 1, the {\it Gaia} G, BP, and RP photometry is converted into the UNIONS $griz$ photometry with a series of color terms. The source density is sufficient to independently calibrate each SEF. In Phase 2, the spectra from {\it Gaia} DR3 are multiplied by the UNIONS filter bandpasses to produce in-field standards for each SEF. The source density of {\it Gaia} sources with spectra is lower; therefore the {\it Gaia} calibration sources are used to bootstrap a higher density catalog derived from the SEFs themselves. OU-EXT then stacks the images to produce co-adds.

OU-MER \cite[Organizational Unit: Merge,][]{oumer} takes the coadds produced by OU-EXT and merges them with the {\it Euclid} data. Forced photometry using SourceXtractor++\citep{2020ASPC..527..461B} and TPHOT \citep{2015A&A...582A..15M} is generated using the {\it Euclid} data as a prior. Images from the space-based {\it Euclid} telescope are significantly higher resolution than the ground-based UNIONS images. {\it Euclid} can resolve multiple adjacent sources that may be merged in the UNIONS data. Thus, the resulting catalog is optimized specifically for {\it Euclid}. {\it Euclid} data releases will be built incrementally; the catalog in each {\it Euclid} data release will only contain UNIONS data over the {\it Euclid} footprint available at the time of that data release.

\section{Summary}

UNIONS is using CFHT, the Pan-STARRS telescopes, and the Subaru Observatory to obtain $ugriz$ images in a core survey area of  6250~deg$^2$ of the northern sky. The $10\sigma$ point source depth of the data, as measured within a 2$^{\prime\prime}$ aperture, are $[u,g,r,i,z] = [23.7, 24.5, 24.2, 23.8, 23.3]$\,mags. UNIONS is a stand-alone scientific survey in its own right that has already contributed to more than 30 articles in the peer-reviewed literature, from near-field cosmology to the low surface brightness Universe, weak lensing, and cosmology. It is set to become the major ground-based legacy survey for the northern hemisphere for the next decade, and provides an essential northern complement to the static-sky science of the LSST. UNIONS supports the core science mission of the {\it Euclid} space mission by providing the data necessary at $\delta \ge 15^\circ$ for the derivation of photometric redshifts in the north as well as other science that requires multi-band photometry. 


\section*{Acknowledgments}

This work is based on data obtained as part of the Canada-France Imaging Survey, a CFHT large program of the National Research Council of Canada and the French Centre National de la Recherche Scientifique. Based on observations obtained with MegaPrime/MegaCam, a joint project of CFHT and CEA Saclay, at the Canada-France-Hawaii Telescope (CFHT), which is operated by the National Research Council (NRC) of Canada, the Institut National des Science de l’Univers (INSU) of the Centre National de la Recherche Scientifique (CNRS) of France, and the University of Hawai'i. 

The Pan-STARRS1 Survey (PS1) and the PS1 public science archive have been made possible through contributions by the Institute for Astronomy, the University of Hawai'i, the Pan-STARRS Project Office, the Max Planck Society and its participating institutes, the Max Planck Institute for Astronomy, Heidelberg, and the Max Planck Institute for Extraterrestrial Physics, Garching, The Johns Hopkins University, Durham University, the University of Edinburgh, the Queen’s University Belfast, the Harvard-Smithsonian Center for Astrophysics, the Las Cumbres Observatory Global Telescope Network Incorporated, the National Central University of Taiwan, the Space Telescope Science Institute, the National Aeronautics and Space Administration under grant No. NNX08AR22G issued through the Planetary Science Division of the NASA Science Mission Directorate, the National Science Foundation grant No. AST-1238877, the University of Maryland, Eotvos Lorand University (ELTE), the Los Alamos National Laboratory, and the Gordon and Betty Moore Foundation.

Pan-STARRS is a project of the Institute for Astronomy of the University of Hawaii, and is supported by the NASA SSO Near Earth Observation Program under grants 80NSSC18K0971, NNX14AM74G, NNX12AR65G, NNX13AQ47G, NNX08AR22G, 80NSSC21K1572, and by the State of Hawaii.

Based on data collected at the Subaru Telescope and retrieved from the HSC data archive system, which is operated by Subaru Telescope and Astronomy Data Center at National Astronomical Observatory of Japan.

The Hyper Suprime-Cam (HSC) collaboration includes the astronomical communities of Japan and Taiwan, and Princeton University. The HSC instrumentation and software were developed by the National Astronomical Observatory of Japan (NAOJ), the Kavli Institute for the Physics and Mathematics of the Universe (Kavli IPMU), the University of Tokyo, the High Energy Accelerator Research Organization (KEK), the Academia Sinica Institute for Astronomy and Astrophysics in Taiwan (ASIAA), and Princeton University. Funding was contributed by the FIRST program from Japanese Cabinet Office, the Ministry of Education, Culture, Sports, Science and Technology (MEXT), the Japan Society for the Promotion of Science (JSPS), Japan Science and Technology Agency (JST), the Toray Science Foundation, NAOJ, Kavli IPMU, KEK, ASIAA, and Princeton University. 

This paper makes use of LSST Science Pipelines software developed by the Vera C. Rubin Observatory. We thank the Rubin C. Observatory for making their code available as free software at \url{https://pipelines.lsst.io}.

This research used the facilities of the Canadian Astronomy Data Centre operated by the National Research Council of Canada with the support of the Canadian Space Agency. 

We are honored and grateful for the opportunity of observing the Universe from Maunakea and Haleakala, which both have cultural, historical and natural significance in Hawaii. 

This work has made use of data from the European Space Agency (ESA) mission
{\it Gaia} (\url{https://www.cosmos.esa.int/gaia}), processed by the {\it Gaia}
Data Processing and Analysis Consortium (DPAC,
\url{https://www.cosmos.esa.int/web/gaia/dpac/consortium}). Funding for the DPAC
has been provided by national institutions, in particular the institutions
participating in the {\it Gaia} Multilateral Agreement.

This work was supported by JSPS KAKENHI Grant Numbers JP24K00684 (HF), JP20H05856 (SM), JP23K22537 (YT), JP22K21349 (YF), JP23K13149 (YF), JP24K22894 (MO). This work was supported by JSPS Core-to-Core Program (grant number: JPJSCCA20210003).

MJH acknowledges support from NSERC through a Discovery Grant 

This work was supported by the TITAN ERA Chair project (contract no. 101086741) within the Horizon Europe Framework Program of the European Commission.

CS acknowledges the support of a NSERC Postdoctoral Fellowship and a CITA National Fellowship.

H. Hildebrandt is supported by a DFG Heisenberg grant (Hi 1495/5-1), the DFG Collaborative Research Center SFB1491, an ERC Consolidator Grant (No. 770935), and the DLR project 50QE2305.

OM is grateful to the Swiss National Science Foundation for financial support under the grant number  PZ00P2\_202104. 

DC wishes to acknowledge funding from the Harding Distinguished Postgraduate Scholarship.

MB acknowledge funding through VIDI grant “Pushing Galactic Archaeology to its limits" (with project number VI.Vidi.193.093) which is funded by the Dutch Research Council (NWO).

LCP acknowledges support from the Natural Sciences and Engineering Research Council of Canada

N.H. is grateful to the Swiss National Science Foundation for financial support under the grant number PZ00P2\_202104.

LB is supported by the PRIN 2022 project EMC2 - Euclid Mission Cluster Cosmology: unlock the full cosmological utility of the Euclid photometric cluster catalog (code no. J53D23001620006).

ES acknowledges funding through VIDI grant "Pushing Galactic Archaeology to its limits" (with project number VI.Vidi.193.093) which is funded by the Dutch Research Council (NWO). This research has been partially funded from a Spinoza award by NWO (SPI 78-411).

This research used the Canadian Advanced Network For Astronomy Research (CANFAR) operated in partnership by the Canadian Astronomy Data Centre and The Digital Research Alliance of Canada with support from the National Research Council of Canada, the Canadian Space Agency, CANARIE and the Canadian Foundation for Innovation.
\bibliographystyle{aasjournal}
\bibliography{main}{}

\end{document}